\documentclass[fleqn,usenatbib]{mnras}

\usepackage{newtxtext,newtxmath}
\usepackage{soul,xcolor}

\usepackage[T1]{fontenc}
\usepackage{ae,aecompl}

\usepackage{graphicx}	% Including figure files
\usepackage{amsmath}	% Advanced maths commands
\usepackage{amssymb}	% Extra maths symbols
\usepackage{xcolor}     
\usepackage{xspace}
\usepackage{multirow}
\usepackage{pdflscape}	% Landscape pages
\usepackage{dblfloatfix}

\def\gsim{\lower.5ex\hbox{\gtsima}} 
\def\lsim{\lower.5ex\hbox{\ltsima}} 
\def\gtsima{$\; \buildrel > \over \sim \;$} 
\def\ltsima{$\; \buildrel < \over \sim \;$} 
\def\gsim{\lower.5ex\hbox{\gtsima}} 
\def\lsim{\lower.5ex\hbox{\ltsima}} 
\def\simgt{\lower.5ex\hbox{\gtsima}} 
\def\simlt{\lower.5ex\hbox{\ltsima}} 

\newcommand{\cii}{[C$\scriptstyle\rm II$]}	
\newcommand{\oiii}{[O$\scriptstyle\rm III$]}	
\newcommand{\loiii}{$L_{\rm [OIII]}$}	
\newcommand{\lcii}{$L_{\rm [CII]}$}
\newcommand{\snr}{SNR}

\newcommand{\lya}{Ly$\alpha$}

\newcommand{\sfr}{M${_\odot}$ yr$^{-1}$}	
\newcommand{\msun}{M${_\odot}$}	
\newcommand{\zsun}{Z${_\odot}$}	
\newcommand{\lsun}{L${_\odot}$}	
\newcommand{\kms}{${\rm km~s^{-1}}$}	

\newcommand{\bdf}{BDF-3299}	
\newcommand{\macs}{MACSJJ149-JD1}	
\newcommand{\yd}{A2744-YD4}	
\newcommand{\sxdf}{SXDF-NB1006-2}

\title[Missing \cii\ emission]{Missing [CII] emission from early galaxies}

\author[Carniani et al.]{S.~Carniani$^{1}$\thanks{E-mail: stefano.carniani@sns.it},
A.~Ferrara$^{1}$,
R.~Maiolino$^{2,3}$,
M.~Castellano$^{4}$,
S.~Gallerani$^{1}$,\newauthor
A.~Fontana$^{4}$
M.~Kohandel$^{1}$,
A.~Lupi$^{1}$,
A.~Pallottini$^{1}$, 
L.~Pentericci$^{4}$, \newauthor
L.~Vallini$^{5}$,
and E.~Vanzella$^{6}$.
\\
$^{1}$ Scuola Normale Superiore, Piazza dei Cavalieri 7, I-56126 Pisa, Italy \\
$^{2}$ Institute of Astronomy, University of Cambridge, Madingley Road, Cambridge CB3 0HA, UK \\
$^{3}$ Kavli Institute for Cosmology, University of Cambridge, Madingley Road, Cambridge CB3 0HA, UK \\
$^{4}$ INAF - Osservatorio Astronomico di Roma, Via Frascati 33, I-00078 Monte Porzio Catone (RM), Italy \\
$^{5}$ Leiden Observatory, Leiden University, PO Box 9500, 2300 RA Leiden, The Netherlands \\
$^{6}$ INAF - Osservatorio di Astrofisica e Scienza dello Spazio, via Gobetti 93/3, 40129 Bologna, Italy
}
\date{Accepted XXX. Received YYY; in original form ZZZ}

\pubyear{2020}

\begin{document}
\label{firstpage}
\pagerange{\pageref{firstpage}--\pageref{lastpage}}
\maketitle

\begin{abstract}
ALMA observations have revealed that \cii 158$\mu$m line emission in high-$z$ galaxies is $\approx 2-3 \times$ more extended than the UV continuum emission. Here we explore whether surface brightness dimming (SBD) of the \cii\ line  is responsible for  the reported \cii\ deficit, and the large $L_{\rm [OIII]}/L_{\rm [CII]}$ luminosity ratio measured in early galaxies. We first analyse archival ALMA images of nine $z>6$ galaxies observed  in both \cii\ and \oiii. After performing several {\it uv}-tapering experiments to optimize the identification of extended line emission, we detect \cii\ emission in the whole sample, with an extent systematically larger than the \oiii\ emission. Next, we use interferometric simulations to study  the effect of SBD on the line luminosity estimate. About 40\% of the extended \cii\ component might be missed at an angular resolution of 0.8\arcsec, implying that $L_{\rm [CII]}$ is underestimated by a factor $\approx2$ in data at low ($<7$) signal-to-noise ratio. 
By combining these results, we conclude that $L_{\rm [CII]}$ of $z>6$ galaxies lies, on average, slightly below the local $L_{\rm [CII]}-SFR$ relation ($\Delta^{z=6-9}=-0.07\pm0.3$), but within the intrinsic dispersion of the relation.
SBD correction also yields $L_{\rm [OIII]}/L_{\rm [CII]}<10$, i.e. more in line with current hydrodynamical simulations. 
\end{abstract}


%%%%%%%%%%%%%%%%%%%%%%%%%%%%%%%%%%%%%%%%%%%%%%%%%%

%%%%%%%%%%%%%%%%% BODY OF PAPER %%%%%%%%%%%%%%%%%%

\section{Introduction}

In the last decade several deep multi-band imaging surveys have identified a large number of galaxies at early epochs. The emerging picture indicates that the cosmic period called Epoch of Reionization (EoR; $6 < z < 10$) is crucial in determining the assembly history of normal star-forming galaxies. Therefore, the characterisation of the interstellar medium (ISM) and star formation processes in galaxies at $z > 6$ is fundamental to understand the early phases of galaxy formation and evolution \citep{Dayal:2018}.

The advent of the Atacama Large Millimetre Array (ALMA) has enabled the first studies of the ISM in $z>4$  ``normal'' star-forming galaxies with star-formation rates (SFRs) $<100$~\sfr,  comparable to those observed in low-$z$ main-sequence galaxies \citep[][]{Ouchi:2013, Ota:2014, Maiolino:2015, Capak:2015, Knudsen:2016,  Inoue:2016,  Pentericci:2016, Bradac:2017,  Matthee:2017, Carniani:2017, Carniani:2018a, Hashimoto:2018, Matthee:2019, Laporte:2019, Harikane:2019, Le-Fevre:2019, Bakx:2020}.
ALMA observations of rest-frame far-infrared (FIR) continuum emission and FIR fine-structure lines, such as \cii($\lambda158\mu$m) and \oiii($\lambda88\mu$m), can  provide direct measurements of dust mass and temperature \citep[e.g][]{behrens:2018,Bakx:2020}, molecular gas content \citep{zanella:2018, Pallottini:2017}, metallicity \citep[e.g.][]{Vallini:2015, Olsen:2017}, SFR \citep[e.g.][]{De-Looze:2014, Herrera-Camus:2015, Carniani:2018a, Schaerer:2020}, gas density, and ionisation parameter \citep{Ferrara:2019, Harikane:2019}. 

The ALMA Large Program to INvestigate \cii\ at Early times (ALPINE) survey has provided the first large sample of star-forming galaxies at $4<z<6$ \citep{Le-Fevre:2019}. 
The sample includes 122 galaxies with SFR$>10$~\sfr\ and stellar mass $10^9$~\msun~$<M_{\star}< 10^{10.5}$~\msun. To optimise the detectability of diffuse emission, all galaxies have been observed in ALMA band~7 by adopting the most compact-array configurations \citep{Bethermin:2020} corresponding to angular resolutions $>1$\arcsec($\approx~6.7$~kpc at $z=4.5$).
The \cii\ line has been detected in 2/3 of the galaxies in the ALPINE survey and, by taking into account the upper limits from non-detections, \citet{Schaerer:2020} have found that the \cii\ luminosity (\lcii) scales linearly with SFR, in agreement with the relation observed in the local Universe \citep[e.g.][]{De-Looze:2014}. This  indicates no (or little) evolution of the \lcii-SFR relation over the cosmic time up to $z\sim6$. 

At higher redshifts ($z>6$), millimetre-interferometer observations have unveiled a more complex scenario. A large fraction ($\sim50\%$) of the galaxy population observed in \cii\ is characterized by a multi-component morphology \citep{Matthee:2017, Jones:2017, Carniani:2018} showing spatial offsets between \cii\ emission and the star-forming regions traced by rest-frame UV light \citep{Matthee:2017,Carniani:2017, Carniani:2018a, Matthee:2019}.
By taking into account such multi-component nature, some studies have shown that the \lcii-SFR relation at early epochs seems to be fully consistent with the local relation \citep{Matthee:2017, Carniani:2018, Matthee:2019}, but its intrinsic scatter is two times larger than observed locally \citep{Carniani:2018a}.
Nevertheless, some studies have highlighted  that   galaxies with  $SFR<30-50$~\sfr\ and/or $z>8$ are systematically  below the local \lcii-SFR relation \citep{Pentericci:2016,Knudsen:2016, Bradac:2017, Matthee:2019, Laporte:2019}. Even taking into account a larger dispersion, some of these galaxies deviates from the  relation more than $2\sigma$.  Such results may indicate that the \lcii-SFR slope changes at low SFRs \citep{Matthee:2019}, and/or that the relation itself evolves at $z>6$ \citep{Laporte:2019}. 

Among  $z>6$ \cii\ emitters observed with ALMA so far, nine UV-selected galaxies (SFRs~$\lesssim100$~\sfr) and three submillimeter galaxies (SFRs~$\gtrsim 300$~\sfr)  have been observed in \oiii\ as well \citep{Inoue:2016, Carniani:2017, Laporte:2017,  Walter:2018, Marrone:2018, Tamura:2019, Hashimoto:2019, Harikane:2019}.
The \oiii\ line has been detected in all galaxies and the reported \oiii/\cii\ luminosity ratios spans a range between 1 and 20, which is  systematically higher than the average  line ratio observed in local star-burst and metal-poor dwarf galaxies \citep{Harikane:2019}.
The observed \oiii/\cii\ ratios are also in tension with most of current cosmological and zoom-in simulations, which struggles to predict FIR luminosity ratios $>2$~\citep[][]{Pallottini:2017a, Olsen:2017, Katz:2017, Katz:2019, Lupi:2020}.
%
%Indeed, while some of these works find high {\it surface density} ratio (i.e. $\Sigma_{\rm [OIII]}/\Sigma_{\rm [CII]} \simeq 10$) in some regions mostly located in the central parts of the simulated galaxies, the total luminosity ratio is typically $\simlt 1$ \citep{Pallottini:2019}.
%
On the other hand, cosmological hydrodynamic simulations by  \citet{Arata:2020} have suggested that high \oiii/\cii\ luminosity ratios, more in line with ALMA observations,  can occur during starburst phases.

The origin of the relatively low (high) \cii~(\oiii) luminosities reported in $z>6$ normal\footnote{galaxies with ${\rm SFR\lesssim100}$~\sfr~that represent the bulk of galaxy population.} star-forming galaxies is still debated. 
Several studies have speculated on different explanations to reproduce current interferometric observations.
\citet{Vallini:2015} find that the \cii\ luminosity decreases with decreasing  gas metallicity and, therefore, the  \cii\ deficit  may indicate that these galaxies are very metal-poor systems. However, recent simulations and theoretical models have shown that gas metallicity, unless very low (e.g. $Z < 0.1 \,$\zsun),  plays a sub-dominant role in shaping the \cii-SFR relation \citep{Ferrara:2019, Pallottini:2019, Lupi:2020}. Upward deviations with respect to the Kennicutt-Schmidt relation due to a starburst phase, could strongly depress \cii\ emission.  This is because the associated strong interstellar radiation field depletes the C$^{+}$ ion abundance by turning C into higher ionization states \citep{Ferrara:2019}. For the same reason, the abundance of \oiii\ in the ionized layer is enhanced. The combination of the two effects boosts the \oiii/\cii\ luminosity ratio \citep{Harikane:2019,Arata:2020}. 
The low \cii\ emission could also be associated to a low (0-10\%) PDR covering fraction due to the compact size of high-$z$ galaxies or galactic outflows \citep{Harikane:2019}. The latter scenario seems to be also supported by recent observational evidences revealing outflowing gas in star-forming galaxies at $z=4-6$ \citep{Gallerani:2018, Fujimoto:2019,Sugahara:2019, Ginolfi:2020}.
Finally, \cite{Kohandel:2019} discuss that the line width of the FIR  line, and thus disc inclination, may be responsible of the non-detections. In fact, at a fixed line luminosity and spectral resolution, narrower emission lines  easily push the peak flux above the detection limit with respect to  broader lines.  If the \oiii\ and \cii\ had different line profile, the FIR line with narrower line width would be easily detectable.

Another possible scenario for the  \cii\ deficit is that current FIR luminosity measurements  suffer from flux losses due the spatially-extended emission of the carbon line. For example \cite{Carniani:2017} show that  about 70\% of the diffuse \cii\ emission of BDF-3299, a star-forming galaxy at $z=7$, is missed in ALMA observations  with angular resolution ($\theta_{\rm beam}$) of $0.3\arcsec$, while the total emission is   recovered in the datasets with $\theta_{\rm beam} = 0.6\arcsec$. 
%In some cases the flux 
%
Another similar case is Himiko, a  \lya\ emitter a $z\sim6$. The flux losses due to the surface brightness dimming led to a non-detection of the FIR line in first ALMA  project. The line was detected successively in a later ALMA program with similar sensitivity but lower angular-resolution  \citep{Ouchi:2013, Carniani:2018}.
It is therefore  fundamental to quantify the effect of  angular resolution on the FIR line luminosity measurements in order to investigate in details the \cii-SFR relation and the \oiii/\cii\ relation in the EoR.

Here we thus focus on the spatial extension of the FIR lines and the impact of surface brightness dimming (SBD) on the line detection and flux measurements at $z>6$.  We re-analyse the  ALMA data of all those UV-selected $z=6-9$ star-forming galaxies observed in \cii\ and \oiii\ in order to verify the robustness of some \cii\ non-detection and  compare the extent of the two FIR emission lines. We then compare the observations with simulations   to verify whether  or not the \cii\ luminosity could be underestimated because the line flux is spatially resolved out. Finally we investigate the \lcii-SFR and \loiii/\lcii\ line ratios at $z>6$ by taking into account the SBD effect. 
The paper is organised as follows. ALMA observations and data reduction are presented in  Sec.~\ref{sec:obs}, while their analysis is discussed in Sec.~\ref{sec:alma_data}. In Sec.~\ref{sec:missing}, we discuss mock ALMA observations in order to  investigate the dependence of luminosity measurements\footnote{We adopt the cosmological parameters from \cite{Planck-Collaboration:2015}: H$_0$ = 67.7 km s$^{-1}$ Mpc$^{-1}$, $\Omega_{\rm m}$ = 0.308 and  $\Omega_{\rm \Lambda}$ = 0.70, according to which 1\arcsec\ at $z = 6$ corresponds to a proper distance of 5.84~kpc. SFR estimates have been calculated by using the relations reported in \cite{Kennicutt:2012}, assuming a \citet{kroupa:01} IMF.} on both angular resolution and sensitivity; we also compare the simulations with real data to assess our results. We then discuss the implications on the \lcii-SFR relation and \oiii/\cii\ luminosity ratio in Sec.s~\ref{sec:relation} and ~\ref{sec:line_ratio}, respectively. Finally, we summarise and draw our conclusions in Sec.~\ref{sec:conclusion}.

\section{Observations and data analysis}\label{sec:obs}

We have retrieved ALMA archival data\footnote{We note that for \sxdf\ and \bdf\ we use additional public datasets (2016.A.00018.S and 2016.1.00856.S) that were not included in previous studies \citep{Inoue:2016, Carniani:2017}.} for galaxies in the literature that have been observed in both \cii\ and \oiii. The list of the sources is shown in Table~\ref{tab:summary}.

The observations have been calibrated with the pipeline script delivered with the raw data from the archive, and by using the Common Astronomy Software Applications package (\verb'CASA'; \citealt{McMullin:2007}). We have used the appropriate package version for each target, as indicated in the pipeline scripts. The final datacube for each target has been generated with the \verb'tclean' task by selecting a pixel scale as large as 1/5 of the ALMA beam. We have used a natural weighting that returns the best surface brightness sensitivity.
We have not  performed any spectral rebinning, i.e. preserving the original spectral resolution of the raw data. The sensitivities of the final cubes (Table~\ref{tab:appendix}) are consistent with those reported by previous works.
For the non-detections, we  have also produced  cubes with lower angular resolution by performing different {\it uv}-tapering, from 0.2\arcsec\ to 2\arcsec.  As explained in Sec.~\ref{sec:alma_data}, decreasing the angular resolution of the final images is crucial to detect faint, extended emission. The {\it uv}-tapering procedure has enabled us to detect the \cii\ line in those galaxies in which previous works quoted a non-detection.

We have constructed the \cii\ flux map with the \verb'CASA' task \verb'immoments' by integrating the channels of the line in the final datacube. 
The integrated flux density has been estimated from the region that encompasses the 2$\sigma$ contours around the peak in the flux map. In those galaxies where the area of the 2$\sigma$-contour region is smaller than 2$\times$ ALMA-beam area, we measure the integrated flux density from a circular  aperture with diameter as large as 1.5$\times$ the major-axis of the ALMA beam and centred at the emission peak of the flux map. We have then calculated the uncertainty on the flux density rescaling the noise by the square root of the number of independent beams in the selected region.

We have also estimated the  extent of the FIR line emission by performing a 2D-Gaussian fitting of the \cii\ flux map with the \verb'CASA' task \verb'imfit'. In addition to the image plane analysis, we have also performed the size measurements on the {\it uv} plane by collapsing the spectral channels around the line peak and following the procedure explained in \citep{Carniani:2019}, which adopts the GALARIO package by \cite{Tazzari:2018}. The two measurements are in agreement within the errors.

ALMA observations properties and all measurements   are reported in Tables~\ref{tab:summary} and ~\ref{tab:appendix}.  
For those galaxies in which
our results are consistent with those reported in previous
studies we list the measurements of the primary works. In particular, the results of our analysis differ from previous works only in some [CII] data, as discussed in Sec. 3.

In the paper we have also used rest-frame UV images from  Hubble Space Telescope ({\it HST}) and UK Infrared Telescope  (UKIRT) images to compare the location of the \cii\ emission with that of the UV region. The relative astrometry of ALMA and {\it HST} images have been calibrated by matching ALMA calibrator and foreground sources (if any) to the GAIA Data Release 1 catalogue \citep{Gaia-Collaboration:2016}.

\section{Individual targets: results}\label{sec:alma_data}

In this Section, we present the results from our ALMA data analysis for the individual targets listed in Table~\ref{tab:summary}, along with a comparison with previous findings in the literature.

\begin{table*}
%	\centering
	\caption{List of targets observed with ALMA in \cii\ and \oiii, and their FIR line luminosities. All estimates are corrected for magnification.
	\label{tab:summary}
	}
	\begin{tabular}{lccccc}
		\hline
		    Target	 & $z$ & $L_{\rm [CII]}~{\rm [10^{8}L_{\odot}]}$ & $L_{\rm [OIII]}~{\rm [10^{8}L_{\odot}]}$ & $L_{\rm [OIII]}/L_{\rm [CII]}$ & $(L_{\rm [OIII]}/L_{\rm [CII]})^{\rm corr}$  \\
		    $^{(1)}$ &  $^{(2)}$ &  $^{(3)}$&  $^{(4)}$ & $^{(5)}$ & $^{(6)}$ \\
		\hline
		\hline
		    MACS1149-JD1    & 9.11 & $0.12\pm0.03^{\ast a} (<0.04)^{\dagger}$ & $0.74\pm0.16^{b}$ & $6.2\pm2.0$  & $ 4.2\pm1.4$  \\
		    A2744-YD4       & 8.38 & $0.18\pm0.06^{\ast a} (<0.2)^{\dagger}$ & $0.70\pm0.17^{c}$ & $3.9\pm1.6$  & $3.2\pm1.3$  \\
            MACSJ0416-Y1    & 8.31 & $1.4\pm0.2^{\ast d}$               & $12\pm3^{e}$ &  $9\pm2$  & $8\pm2$  \\
            SXDF-NB1006-2   & 7.21  & $1.7\pm0.4^{a} (<0.8)^{\dagger}$       & $9.8\pm2.2^{f}$ & $ 5.8\pm1.9$  & $4.3\pm1.4$   \\
            B14-65666       & 7.16 & $11.0\pm1.4^{g}$              & $34\pm4^{g}$  & $   3.1\pm0.5$  & $2.8\pm0.5$  \\
            BDF-3329        & 7.11 & $0.67\pm0.09^{a}$             & $1.8\pm0.2^{i}$ & $  2.7\pm0.5$ & $1.8\pm0.5$ \\
            J0217           & 6.20 & $14\pm2^{j}$                  & $85\pm2^{j}$ & $   6.1\pm0.9$ & $5.7\pm0.9$   \\
            J0235           & 6.09 & $4.3\pm0.7^{j}$               & $38\pm3^{j}$  & $   8.8\pm1.6$  & $ 7.8\pm1.4$  \\
            J1211           & 6.03 & $14\pm1^{j}$                  & $48\pm7^{j}$  & $   3.4\pm0.6$  & $ 3.4\pm0.6$ 	 \\
		\hline
		\hline
	\end{tabular}	
		
	{\bf Columns.} 
	(1) Name of the target. 
	(2) redshift. 
	(3,4) Observed \cii\ and \oiii\ luminosities. 
	(5) \oiii-to-\cii\ luminosity ratio. 
	(6) \oiii-to-\cii\ luminosity ratio corrected for the SBD effect (see text).
	{\bf Notes.} 
	$^\ast$ Lensed galaxies. We assume a magnification factor $\mu=10, 2,$~and~$1.4$ for \macs,\yd, and MACSJ0416-Y1, respectively.
	$\dagger$ previous upper limits by \citet{Inoue:2016} and \citet{Laporte:2019}.
	% and  report upper limits on \lcii\ of $4\times10^{6}$~\lsun\ and  $2\times10^{7}$~\lsun\ for \macs\ and \yd.  For \sxdf\ \citet{Inoue:2016} find an upper limit on \cii\ of $8.3\times10^7$~\lsun.
	%
	{\bf References}: $^a$ this work;
	$^b$ \citet{Hashimoto:2019}; 
	$^c$ \citet{Laporte:2017}; 
	$^d$ \citet{Bakx:2020}; 
	$^e$ \citet{Tamura:2019};
	$^f$ \citet{Inoue:2016};   
	$^g$ \citet{Hashimoto:2019}; 
	$^h$ \citet{Maiolino:2015};  
	$^i$ \citet{Carniani:2017}; 
	$^j$ \citet{Harikane:2019}.
\end{table*}

\subsection{\sxdf, \yd, and \macs}

We focus initially on \sxdf\, and the two lensed galaxies \yd\, and \macs. In these sources, previous studies have detected  the \oiii\ line but reported a non-detection for \cii\ \citep{Inoue:2016, Laporte:2019}. 
These galaxies seem to be characterised by a high FIR line ratio (\loiii/\lcii~$>10$ for \sxdf\ and \macs, and \loiii/\lcii~$>3$ for \yd) that is a few times higher than those observed in the local Universe \citep{De-Looze:2014, Cormier:2015, Harikane:2019} and simulations \citep{Pallottini:2019, Arata:2020}.

Given the results of previous studies,  we have changed the data reduction method to verify if the \cii\ non-detection is due to resolving out of the line emission.
We have thus generated 10 different ALMA cubes for each target by varying the {\it uv}-taper parameter\footnote{The {\it uv}-tapering procedure reduces the angular resolution  by scaling down the weight of the {\it uv}-data from the longer baselines. It thus smooths the final image but at the expense of sensitivity since part of the data are excluded or data usage is non-optimal.} from 0.2\arcsec\ to 2\arcsec\ in steps of 0.2\arcsec. 
We stress that the  {\it uv}-tapering decreases both angular resolution and sensitivity of the final images.  Therefore our approach of analysing images of the same target with different {\it uv}-tapering and, hence, ALMA beams, enables us to find the best  sensitivity-angular resolution combination that optimises the detection of extended emission. 
In each cube, we have then performed a blind-line search within 5\arcsec\ from the location of the targets and from $-1000$ \kms\ to 1000 \kms\ with respect to the \oiii\ redshift (see App.~\ref{app:blind} for details). Finally, we have selected only those detections with a level of confidence\footnote{We adopt the detection threshold used in the ALPINE survey \citep{Bethermin:2020}.} $>3.6\sigma$, and among this final sample we have extracted the line candidate with the highest confidence level.

\begin{figure*}
	\includegraphics[width=2\columnwidth]{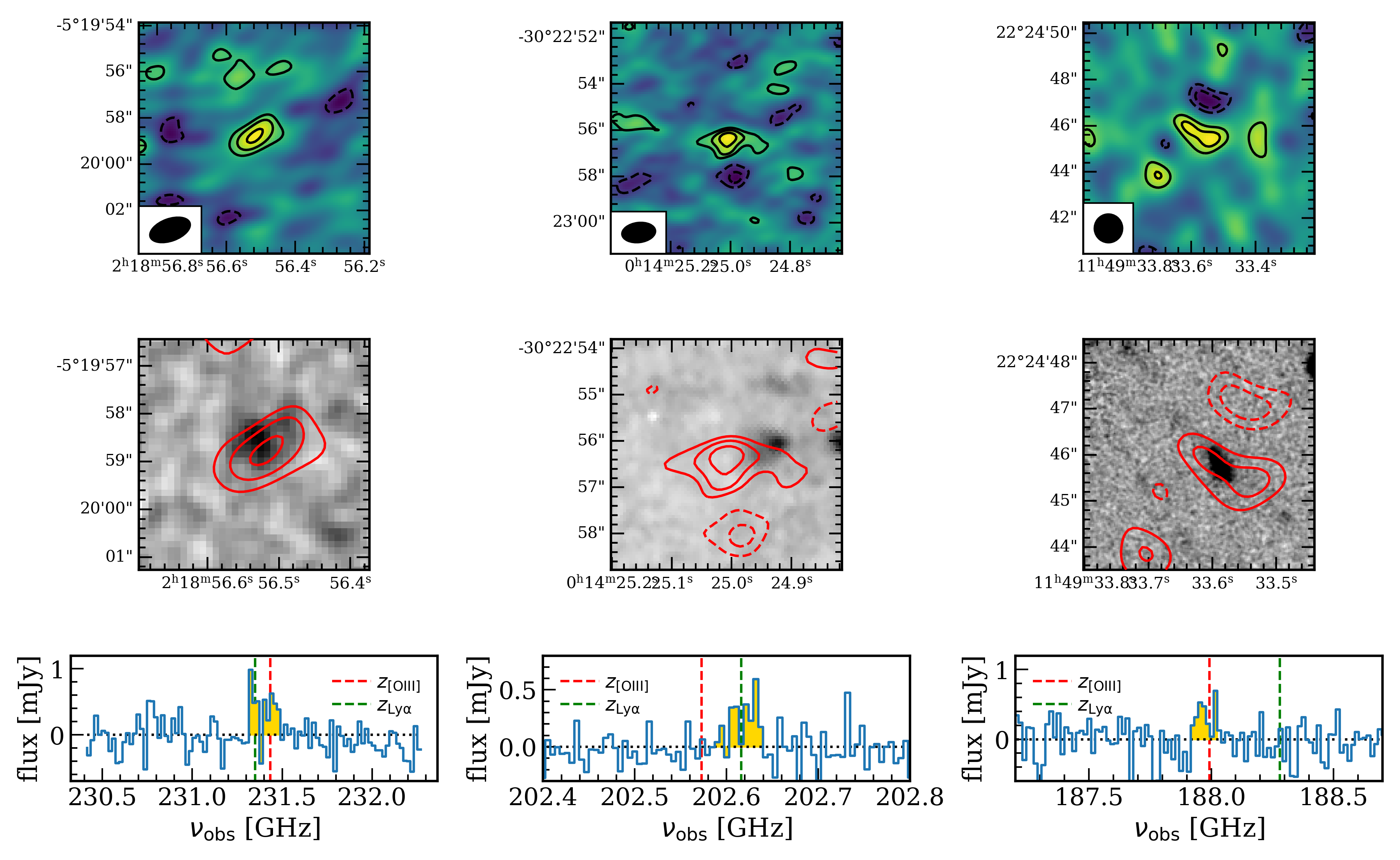}
    \caption{From left to right panels:  \cii\ line detections of \sxdf, \yd, and \macs. In the top row we report the \cii\ maps obtained by collapsing the ALMA datacube over the line width of the detected line. Contours are at level of $\pm2,3,~\rm{and}~4\sigma$. Middle row illustrates zoom-in maps of rest-frame UV emission  and the red contours are the \cii\ emission. The bottom row shows the \cii\ spectra extracted from the 2$\sigma$-contour regions. The \oiii\ and \lya\ redshifts are highlighted with vertical red and green dashed lines, respectively.}
    \label{fig:cii_eor}
\end{figure*}

In all three galaxies we have found a candidate line  with a significance level $>3.8\sigma$ in the integrated {\it uv}-tapered map at the redshift of either \oiii\ or \lya\ line.  
In Fig.~\ref{fig:cii_eor} we show the results of our blind line search procedure.   
Luminosity estimates are reported in Table~\ref{tab:summary} while other measurements are listed in App.~\ref{tab:appendix}. 
The \cii\ detection for both \sxdf\ and \macs\ is co-spatial with the UV emission, while the \cii\ emission in \yd\ is located 0.8\arcsec\ away from the galaxy.  

Given the large ALMA beam used, the spatial offset of \yd\ might be consistent with the astrometric accuracy of ALMA\footnote{Sec. 10 of the ALMA technical handbook}, $\sigma_p$, given by 
\begin{equation}
\sigma_p = 60~{\rm mas}~\left(\frac{100~{\rm GHz}}{\nu_{\rm obs}}\right)\left(~\frac{10~{\rm km}}{b}\right)~\left(\frac{1}{{\rm SNR}}\right)\,,
\end{equation}
where $\nu_{\rm obs}$ is the observing frequency, $b$ is the maximum baseline, and SNR is the signal-to-noise ratio of the source peak. The astrometric uncertainty  for \yd\ ALMA observations is expected to be $\sim0.6$\arcsec. %which is 1.3 times smaller than the spatial offset between the \cii\ emission and the UV region. 
Therefore the location of  the \cii\ and UV regions are   consistent within 1.3$\sigma_p$.
However, we also note that \cite{Laporte:2019} report a spatial offset between the \oiii\ and dust continuum emission as well.
The situation of \yd\ could thus resemble what observed in \bdf, where the \cii, \oiii, and UV emission are tracing different components of the same system with different properties \citep{Carniani:2017}. The spatial offsets could also be associated to either material ejected by galactic outflows or a galaxy merger \citep{Maiolino:2015, Vallini:2015, Pallottini:2017, Katz:2017, Gallerani:2018, Kohandel:2019}. The merger scenario seems to be supported by the fact that the UV images show other group members around \yd\ \citep{Zheng:2014}

\subsection{\bdf}

\cite{Maiolino:2015} and \cite{Carniani:2017} present Cycle-1 and -2 ALMA observations of \bdf, a \lya-emitting galaxy at $z=7.1$ \citep{Vanzella:2011,Castellano:2016}.
ALMA images have revealed a \cii\ emission consistent with the \lya\ redshift  but spatially offset by 0.7\arcsec\ to the optical (UV-rest frame) emission. By using serendipitous sources found in the ALMA field-of-view, the authors conclude that the displacement is not ascribed to astrometric uncertainties.

In  Cycle 4, we were awarded  ALMA time   to  obtain deeper \cii\ observations of \bdf\ with respect to those presented by \citet{Maiolino:2015}. The proposed observations aimed at detecting extended emission around the galaxy, but only 25\% of the proposed program was completed. The achieved sensitivity, $\sigma_{\rm 20 km/s}$\footnote{Sensitivity level in spectral bins of 20~\kms}  =120~$\mu$Jy~beam$^{-1}$, is comparable to that of previous observations \citep{Maiolino:2015, Carniani:2017}.
The program was carried out with a semi-compact array configuration resulting in an angular resolution of $0.6\arcsec\times0.5\arcsec$.

\begin{figure}
    \begin{center}
        \includegraphics[width=\columnwidth]{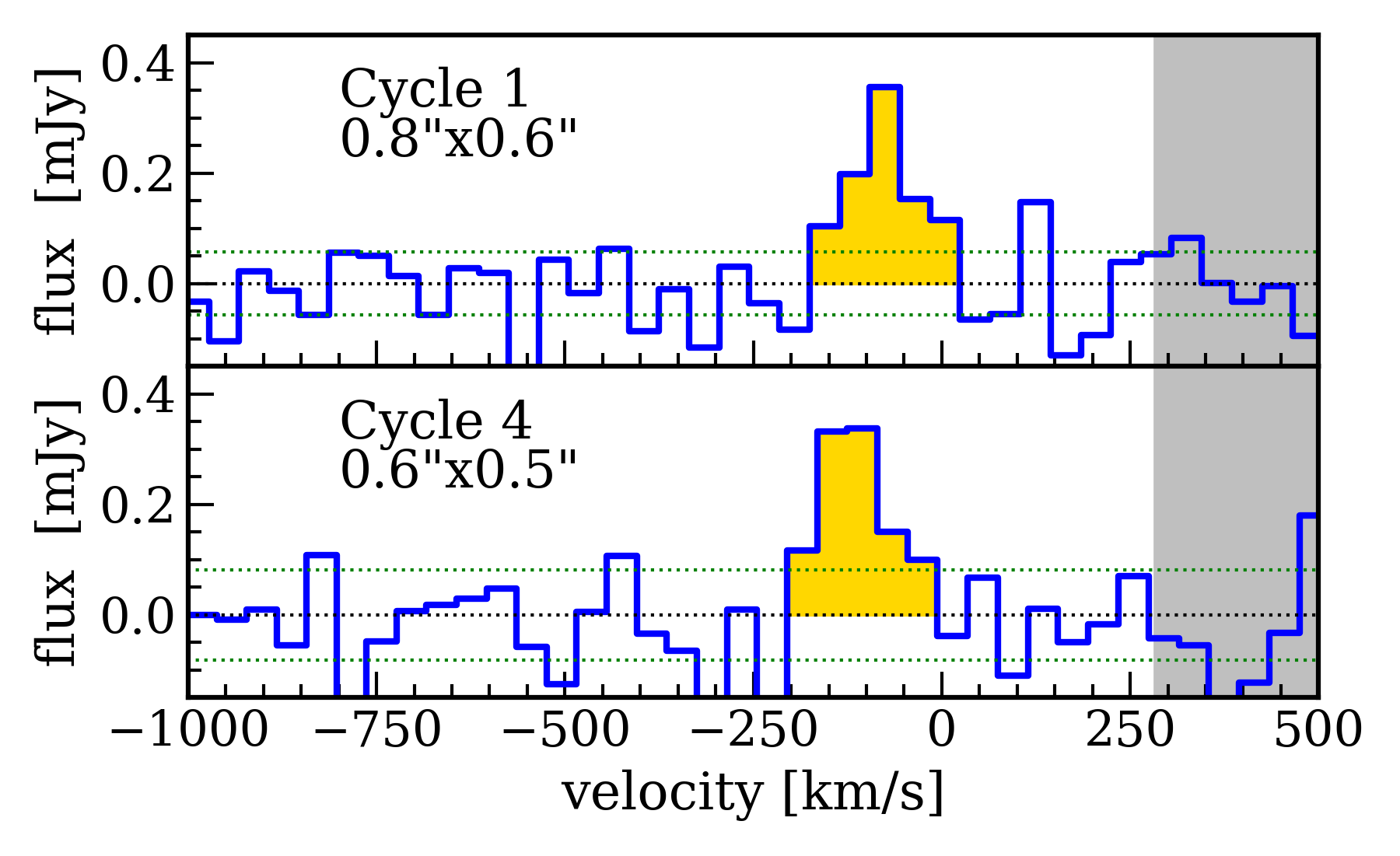}\\
    \end{center}
    \begin{center}
	    \includegraphics[width=\columnwidth]{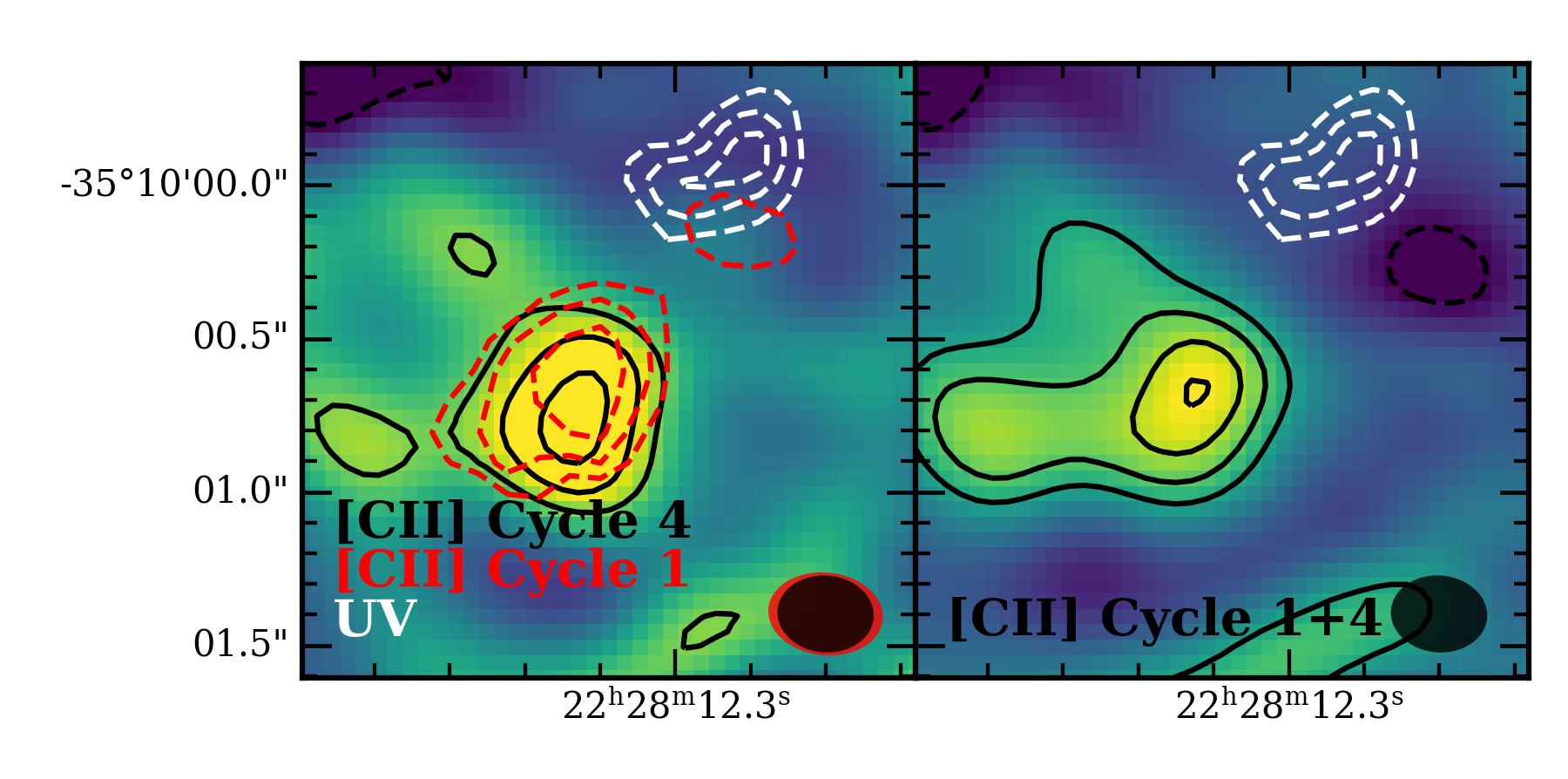}\\
	\end{center}
    \caption{\emph{Top}:  \cii\ spectra of \bdf\ from the old (Cycle 1) and new (Cycle 4) ALMA programs, extracted at the location of \cii\ clump. \emph{Bottom}: Left panel shows the \cii\ map  of \bdf\ produced only from the new Cycle-4 dataset; the right panel illustrates the flux map from the combined dataset (Cycle 1 + Cycle 4) . The black contours show $\pm2$, $\pm3$, $\pm4$, and $\pm5\sigma$.   Red dashed contours trace the Cycle-1 \cii\ map reported by \citet{Maiolino:2015} and contours are at levels 2, 3 and 4 times noise per beam, while white dashed contours correspond to the rest-frame UV emission from {\it HST} observations (\citealt{Castellano:2016}). We report the synthesised ALMA beam in the bottom-right corners.
    \label{fig:bdf_1}
    }
\end{figure}

In the new dataset we have detected the \cii\ line with a level of significance of 5$\sigma$ and spatially offset by $\sim0.7$\arcsec\ ($\sim3.7$~kpc) from the UV emission (Fig.~\ref{fig:bdf_1}). In this case the spatial offset is $3.5$ times larger than the $\sigma_p=0.2\arcsec$ and is unlikely to be related to the astrometric calibration of ALMA dataset.
Both redshift and line width are consistent with the \cii\ properties estimated by \citet{Maiolino:2015} and \citet{Carniani:2017}. This new independent dataset confirms the robustness of the displaced \cii\ detection.

By combing the new and old datasets we have reached a sensitivity of $\sigma_{\rm cont}={\rm 8~\mu Jy~beam^{-1}}$ and  $\sigma_{\rm 20km/s}={\rm 90~\mu Jy~beam^{-1}}$ in the continuum and line map, respectively. Despite the deeper images, we have detected neither the continuum nor \cii\ emission at the location of the UV region.

The bottom panel of Fig.~\ref{fig:bdf_1} shows the flux map of the spatially offset \cii\ emission obtained from the combined dataset. 
The peak of the emission has a significance level of 5.2$\sigma$ and the total integrated flux is $S\Delta v = 52\pm7$ mJy~\kms, which is consistent within the error with previous measurements. We thus infer a  \cii\ luminosity of $(6.7\pm0.9)\times10^7$~\lsun.

\begin{figure}
    \begin{center}
        \includegraphics[width=\columnwidth]{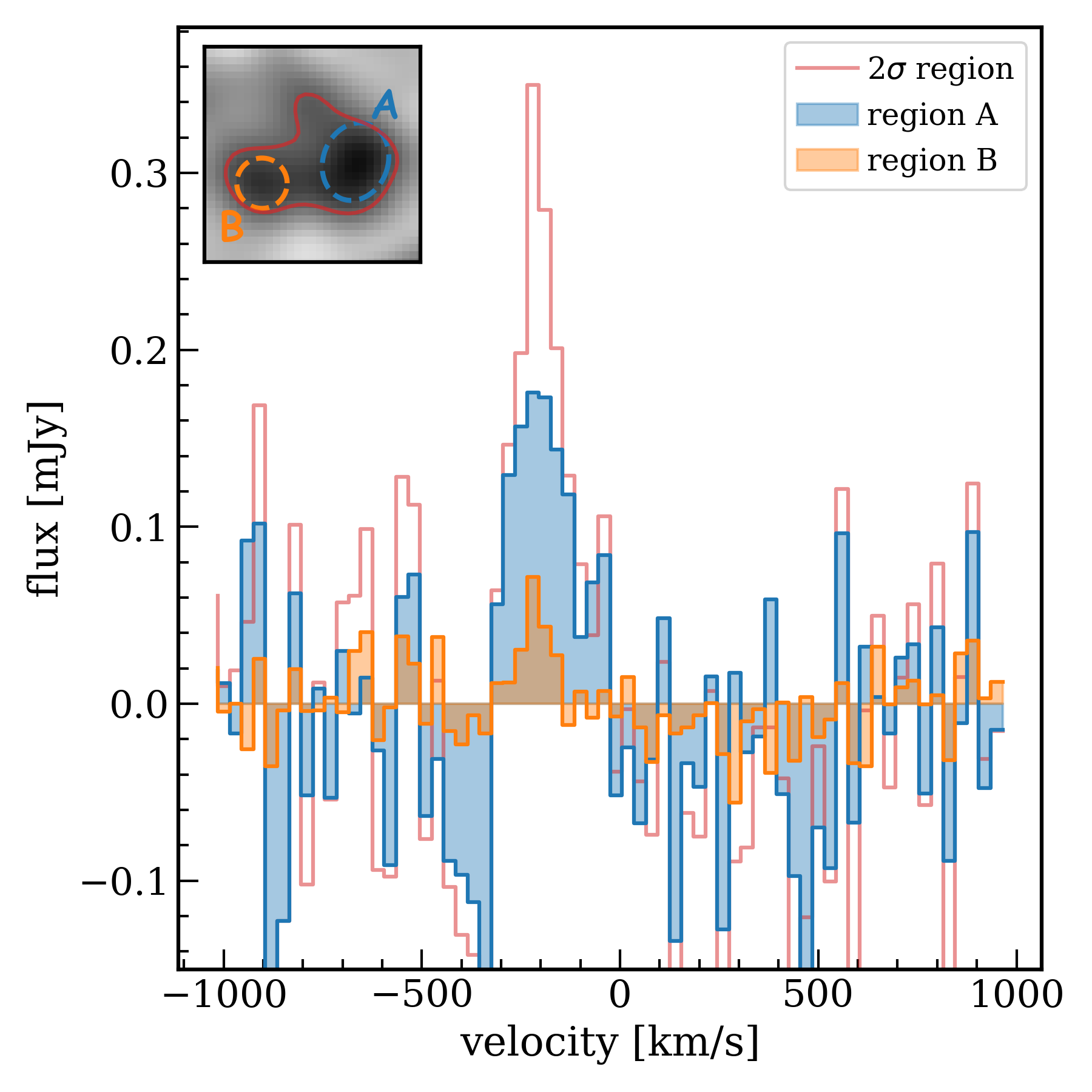}\\
    \end{center}
    \caption{\cii\ spectra of \bdf\ obtained from the combined ALMA dataset. Red line show the spectrum of the total \cii\ emission arising from the 2$\sigma$-contour region (red contour in the inset). The spectra of the individual \cii\ components, A and B, are shown in blue and orange. The top-left inset illustrates the  apertures adopted to extract the spectra of the two \cii\ components.}
    \label{fig:bdf_2}
\end{figure}

In the new map we  notice that the displaced \cii\ emission is more extended with respect to what observed in previous shallower observations. 
The \cii\ morphology of the map can be described by two components, dubbed A and B in Fig.~\ref{fig:bdf_2}. The spectra extracted from the individual regions indicate that the two \cii\ components have similar redshifts but  different line widths ($FWHM_{\rm A}=143$~\kms\ and $FWHM_{\rm B}=80$~\kms). 

We conjecture that the extended emission toward East represents an additional fainter and smaller satellite member  of \bdf\ system. A similar scenario is in agreement with zoom-in simulations \citep{Pallottini:2019}, predicting that high-$z$ systems are surrounded by satellites with SFR~$< 5$~\sfr. These satellites are located within 100~kpc from the main galaxy, and are too faint to be detected in shallow UV  images. However, such satellites could be visible in either deep \cii\ observations as \bdf\ or rest-frame UV {\it HST} observations of strongly lensed systems \citep{Vanzella:2017, Vanzella:2019}. 

Another possibility is that the extended \cii\ emission is tracing metal-enriched circumgalactic medium (CGM) around \bdf. Recent high-$z$ observations have indeed shown that galactic outflows \citep{Gallerani:2018,Fujimoto:2019,Ginolfi:2020} and  gas stripping by tidal interactions \citep{Ginolfi:2020a} might be responsible of the  carbon enrichment on kiloparsec scales. The \cii\ emission on large scale is then excited by the UV radiation of the galaxy \citep{Carniani:2017}. However in this case one would expect to observe a gradient of velocity between the two \cii\ components associated to either outflowing or inflowing kinematics as observed in other high-$z$ systems \citep[e.g.][]{Jones:2017,Ginolfi:2020a}. 
However, distinguishing  between satellites or CGM requires either more sensitive \cii\ images or JWST observations enabling us to identify  the emission (if any) of the stellar population from each individual component.

\subsection{MACSJ0416-Y1, B14-65666, J0235, J1211, and J0217}
For the remaining systems (MACSJ0416-Y1, B14-65666, J0235, J1211, and J0217) for which co-spatial \cii\ and \oiii\ emissions have been already reported in the literature \citep{Tamura:2019, Hashimoto:2019,Harikane:2019, Bakx:2020}, we have found results consistent with previous works. In Table~\ref{tab:summary} we list the \cii\ and \oiii\ luminosities.

\begin{figure}
	\includegraphics[width=\columnwidth]{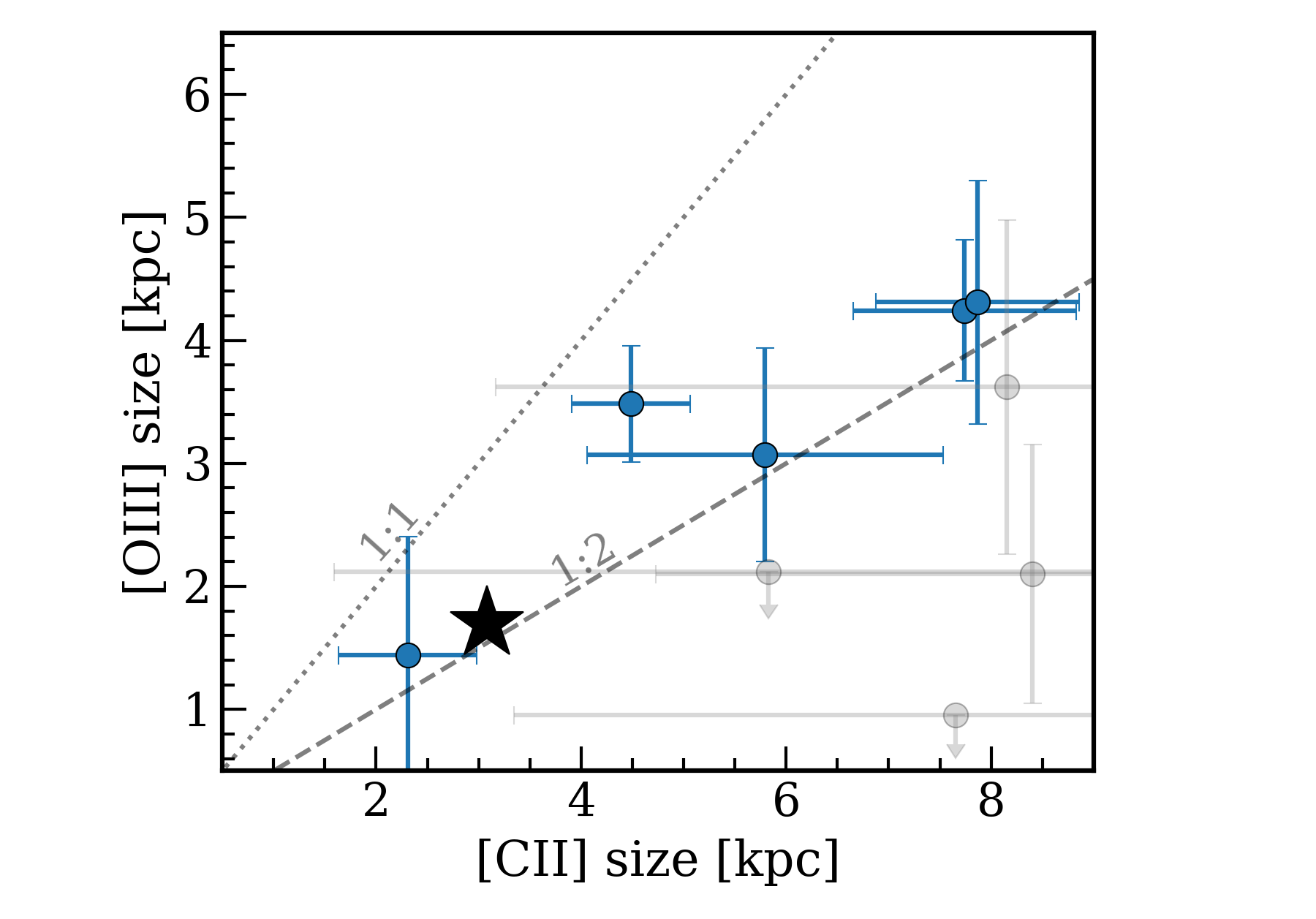}
    \caption{Comparison between \oiii\ and \cii\ emission extension of our sample. The circle blue marks indicate the size measurements of MACSJ0416-Y1, B14-65666, J0235, J1211, and J0217, which are detected with high SNR. The grey circles show the extents of the remaining galaxies identified with a lower SNR (MACS1149-JD1, A2744-YD4, SXDF-NB1006-2, and BDF-3299). The star symbol represents the \oiii\ and \cii\ sizes of Freesia, the most massive galaxy in the \citet{Pallottini:2019} simulations. The dotted and dashed lines indicate 1:1 and 1:2 relation, respectively.} 
    \label{fig:size}
\end{figure}

Since for these galaxies the two FIR lines have been detected with high signal-to-noise ratio (SNR~$>8$), we take advantage of these observations to investigate the extent of the FIR lines, as shown in Fig.~\ref{fig:size} (and Table~\ref{tab:appendix}). We notice that  \cii\ is systematically larger than the \oiii\  and its extent is about two times larger than that of  oxygen line.
This is consistent with  recent simulations by \cite{Pallottini:2019} where the \oiii\ is concentrated in a compact region of 0.85~kpc, while the \cii\ arise from a more extended area with a radius of 1.54~kpc. Simulations and observations suggest that distinct FIR emission lines  trace different regions of the same galaxy, characterised by different metallicity or excitation (ionisation parameter) properties \citep{Carniani:2017, Katz:2017, Pallottini:2019}.

As we will discuss in Sec.~\ref{sec:discussion} there are various scenarios to explain the extended component of the \cii\ line. Here we note that the different extent of \cii\ and \oiii\ emission could dramatically affect the measured \oiii/\cii\ ratios. Indeed, while in images with an intermediate ALMA spatial-resolutions (i.e. $\sim4-5$~kpc) the \oiii\ emission may appear as point-like, \cii\ emission could be spatially resolved and a fraction of the extended emission could be missed due to the low sensitivity. In the next Section we make use of ALMA simulations to quantify this effect in available observations.

\section{\cii\ surface brightness dimming}\label{sec:missing}

We have performed ALMA simulations with different array configurations to estimate how the angular resolution affects the \cii\ and \oiii\ flux measurements. This enables us to assess whether a  fraction of the \cii\ emission might be missed when the FIR line flux is spatially resolved and the sensitivity is too low to recover the total surface brightness.

\subsection{Simulations}\label{sec:simulations}

We have used the \verb'simobserve' task of \verb'CASA' to produce mock interferometric observations of galaxies at $z\sim7$.
As source models, we have used a 2D Gaussian profile for the \cii\ and \oiii\ surface brightness with major-axes\footnote{FWHM of the 2D-Gaussian profile } ($D_{\rm source}$) fixed to 0.8\arcsec~($\approx4.6$~kpc) and 0.45\arcsec~($\approx2.6$~kpc), respectively for the two FIR lines. The assumptions on the $D_{\rm source}$ are based on current ALMA observations \citep[see Fig.~\ref{fig:size} and][]{Carniani:2018a}. We have then assigned to each source a random axis ratio ($0.1 < b/a < 1.0$) and position angle. 

For simplicity, we have assumed \lcii~=~\loiii~=~$5\times10^{8}$~\lsun\ for all mock sources. We have thus set the integrated flux density to 0.43 Jy~\kms\ for the \cii\ sources and 0.21 Jy~\kms\ for the \oiii\ mock targets, and assumed a line width of 200~\kms. 
Once fixed the line properties, we have generated two set of simulations.

For the first set of simulations we have computed 100 synthetic observations for each of the five most compact array configurations at the observed line frequencies of 230~GHz and 410~GHz, which correspond to the redshifted frequencies of the two FIR lines for a galaxy a $z\sim7$.
The on-source exposure time (t$_{\rm exp}$) of each line has been fixed for all simulations in order to obtain the same noise level per ALMA beam independently of array configurations. 
By using the ALMA exposure time calculator (ETC), we have estimated  t$_{\rm exp}$~=~16 minutes for the \cii\ pointings and t$_{\rm exp}$~=~6.5 hours for the \oiii\ pointings. Such t$_{\rm exp}$ enable the detection of the two FIR lines with a level of significance of $\sim$10$\sigma$ if the target is a point-like source. 

In the second set of  mock interferometric data we have run several simulations for each array configuration with different t$_{\rm exp}$.  The emission in the final images has hence different \snr\ depending both on the ALMA beam ($0.4\arcsec<\theta_{\rm beam}<2\arcsec$) and sensitivity ($5~{\rm mJy~km s^{-1}~beam^{-1}} < \sigma < 140~{\rm mJy~km~s^{-1}~beam^{-1}} $). These mock data have allowed us to  investigate the impact of the noise level on the detection of extended emission.

In the resulting mock images of both sets of simulations we have finally measured the FIR line fluxes by adopting the same procedure used in real images and described in  Sec~\ref{sec:obs}.

\subsection{Analysis of mock data}

\subsubsection{Simulations with fixed  exposure times}

\begin{figure*}
	\includegraphics[width=2\columnwidth]{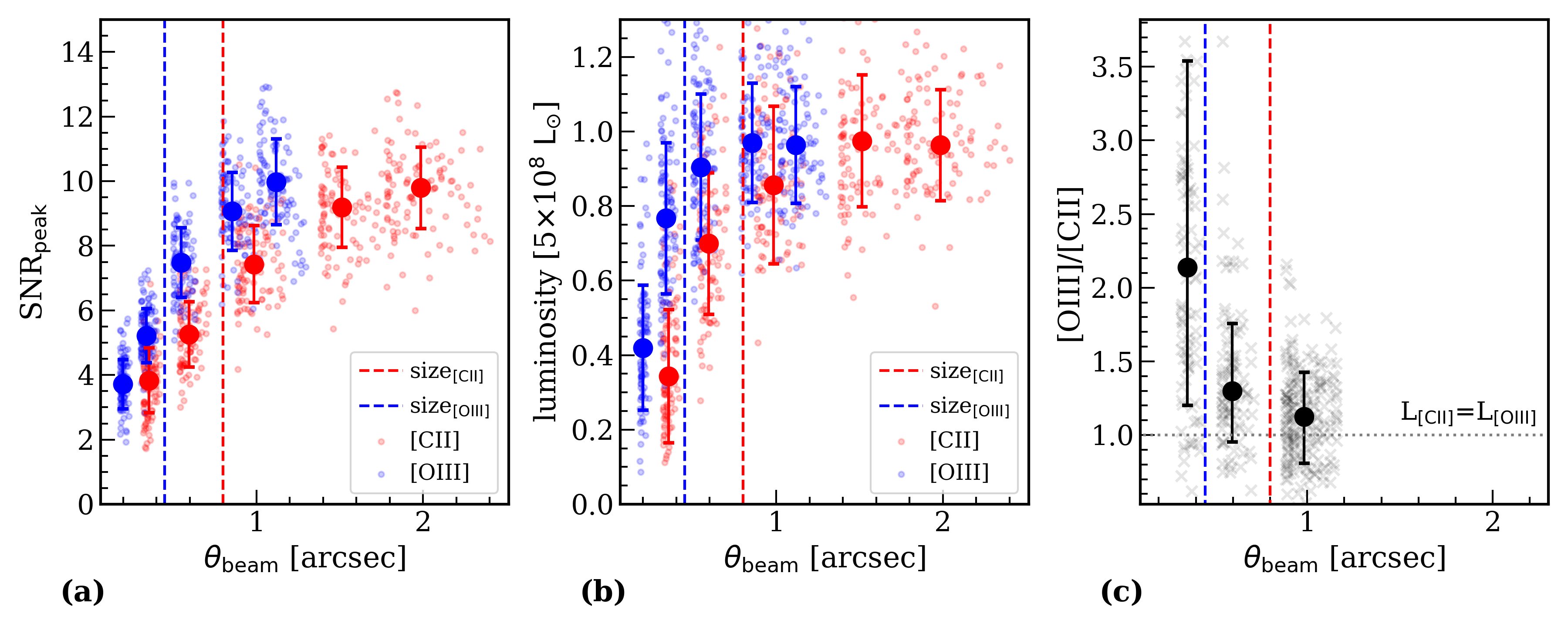}
    \caption{\cii\ (red circles) and \oiii\ (blue circles) synthetic ALMA observations of $z=7$ galaxies with \lcii=\loiii=$5\times10^{8}$~\lsun\ by using different array configurations, but fixing the exposure times: (t$_{\rm exp} = 16$~mins for \cii\ and t$_{\rm exp} = 6.5$~h for \oiii). We have assumed a size for \cii\ and \oiii\ emission of 0.8\arcsec and 0.45\arcsec, respectively. Panels (a) and (b) show the signal-to-noise ratio (\snr) of emission peaks and line luminosities, respectively, as a function of angular resolution. Average values for each array configuration are shown with larger marks.  In panel (c), we report the \oiii/\cii\ luminosity ratios obtained from \oiii\ and \cii\ mock observations with similar angular resolutions.  }
    \label{fig:sim}
\end{figure*}

Let us consider first the set of ALMA simulations obtained with different array configurations but fixed exposure times.  
The left and middle panels of Fig.~\ref{fig:sim} show the \snr\ of the \cii\ and \oiii\ detections and their line luminosities as a function of the angular resolution\footnote{The angular resolution of ALMA image depends on both  the observing frequency ($\nu_{\rm obs}$) and the maximum baselines ($b$) of the adopted array configurations. The FWHM of the ALMA beam is given by  $\theta_{\rm beam}[\arcsec] \approx \frac{76}{b[{\rm km}] \nu_{\rm obs}[{\rm GHz}]}$.} (i.e. ALMA array configuration).
Note that the lowest angular-resolution of the \cii\ images ($\theta_{\rm beam}\sim2\arcsec$) is different from that of \oiii\ ($\theta_{\rm beam}\sim1.2\arcsec$) because of the different frequencies of the two FIR lines.

We notice that the SNR and the measured luminosity of the detections decreases at increasing angular resolution.
At low angular resolutions, when $\theta_{\rm beam}/{\rm D_{\rm source}}\gtrsim1.5$ (i.e.  $\theta_{\rm beam}>1.2\arcsec$ for \cii\ and  $\theta_{\rm beam}>0.7\arcsec$ for \oiii\ observations), both FIR lines are spatially unresolved  and have the maximum SNR ($\sim10$), which is consistent with that returned by the ALMA ETC. 
The measured line luminosities  are consistent with our input values as well.

Moving to higher angular resolutions, the SNR of both FIR lines decreases from 10 to 4. This effect is caused by the surface brightness dimming (SBD) due to the decreasing of the solid angle area, i.e. ALMA beam. 
We also note that in the same range of angular resolutions the measured luminosities fall down to $1.5-2\times10^{8}$~\lsun,  indicating that $\sim60\%-70\%$ of the total luminosity is missed in high-angular resolution (and low SNR) observations.

Despite the similar decreasing trend of the two lines, the SNR and line luminosity of the \cii\ line drops more rapidly with the angular resolution than  the \oiii\ line luminosity.  
The effect of the SBD has indeed a larger impact on the carbon line because the \cii\ emission is more extended than the \oiii. 
For example, at $\theta_{\rm beam}=0.8\arcsec-1\arcsec$ - similar to  \cii\ size - the \oiii\  still appears a point-like source  while the \cii\ emission is spatially resolved and the line luminosity is underestimated by $20\%-40\%$.

As most of current ALMA campaigns targeting \oiii\ and \cii\   in high-$z$ galaxies \citep{Inoue:2016, Carniani:2017, Hashimoto:2019, Harikane:2019, Laporte:2019} have been set to obtain images of both FIR lines with similar angular resolutions, in the right panel of Fig.~\ref{fig:sim} we report the luminosity line ratio obtained from those synthetic observations having a similar ALMA beam for both lines.
At angular resolutions of $\sim1$\arcsec\ we  infer an average line ratio that is 1.15 times larger than the input value. The different extension of the two lines alters the line ratio estimates, in particular yielding an overestimate of the total \oiii/\cii\ luminosity ratio.
At smaller ALMA beams, the ratio estimated from the mock observations is even larger, specifically: 1.35 and 2.1 at 0.6\arcsec\ and 0.4\arcsec, respectively. We  note that this bias is larger than the typical uncertainty associated to the line ratio estimates (Table~\ref{tab:summary})  and should be taken into account when we investigate intensity of FIR lines in the high-$z$ Universe.

\subsubsection{Simulations with different exposure times}

\begin{figure}
	\includegraphics[width=\columnwidth]{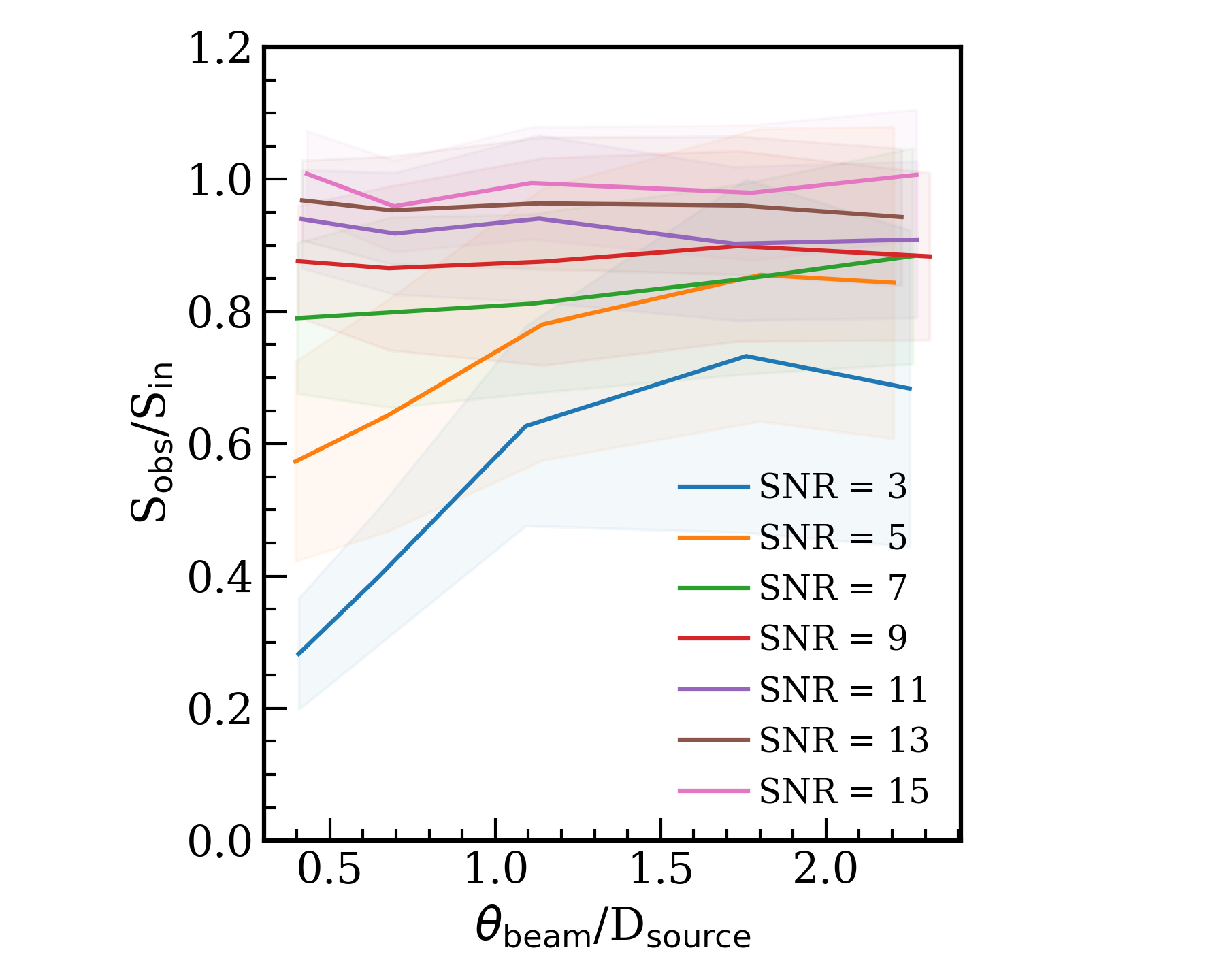}
    \caption{Ratio  between measured and intrinsic fluxes, $S_{\rm obs}/S_{\rm in}$, as a function of ALMA angular resolution normalised to the \cii\ source size, $D_{\rm source}$, and for different signal-to-noise ratios. The flux is extracted from a region that encompasses the 2$\sigma$ contour of the emission. If the area of this region is smaller than 2$\times$ the ALMA beam, we use a circular region with diameter 1.5$\times$ the major-axis of the ALMA beam.}
    \label{fig:corr_vs_sn}
\end{figure}

So far we have assessed the effect of the SBD on the flux measurements at fixed on-source exposure time, i.e. sensitivity per ALMA beam. Now we take advantage of the second set of simulations to determine the bias driven by sensitivity at different angular resolutions. In Fig.~\ref{fig:corr_vs_sn} (and Table~\ref{tab:corr}) we report the ratio between the observed and intrinsic flux of our sources as a function of $\theta_{\rm beam}$ normalised by \cii\ size (D$_{\rm source}$) and for different \snr, i.e. noise levels. 

In synthetic images in which the emission peak is higher than 10$\sigma$ the measurements are in agreement with the input values, independent of the ALMA array configuration; the discrepancy between the observed flux and the model is lower than 10\%, which is of the same order of the noise level and flux calibration uncertainties (i.e. 5-10\%).
At lower signal-to-noise ratios (\snr~$<10$), the discrepancy increases, with the fraction of missed flux depending on the ratio between the angular resolution and the emission extent as well.
In particular, at \snr~=~5  the flux inferred from the mock observations with the most-compact ALMA array is a factor $\sim0.8$ times lower than the input flux, while between 30\% and 60\% of the flux is missed in the extended configurations.

This second set of simulations catches the effect of SBD on the line luminosity measurement for different SNR and angular resolutions. Once validated against real observation, we can  use these results to recover the total FIR line luminosities of high-$z$ galaxies, whose detections have low SNR and are spatially resolved.

\subsection{Comparing mock data with observations}\label{sec:other_obs}

In order to verify the results achieved from our simulations, we have searched in the literature for  \cii-emitting galaxies observed multiple times with different ALMA-array configurations, but similar sensitivity.
We have thus  found that seven of the ten \cii\ emitters reported by \cite{Capak:2015} have been recently observed in the ALPINE survey by using a more compact ALMA-array configuration \citep{Le-Fevre:2019, Bethermin:2020}.
The two datasets, 2012.1.00523.S and 2017.1.00428.L (hereafter C12 and L17), have an angular resolution of $0.7\arcsec$ and $1\arcsec$, respectively.

Among the seven  galaxies in common with the two ALMA programs, we have analysed only three sources, HZ1, HZ3, and HZ4, since the other targets show a multi-component morphology \citep{Carniani:2018} that would lead to a more complex and ambiguous interpretation with respect to current simulations. 
 The two datasets, data calibration, and analysis are presented in App.~\ref{sec:app_hz}.

\begin{figure}
	\includegraphics[width=\columnwidth]{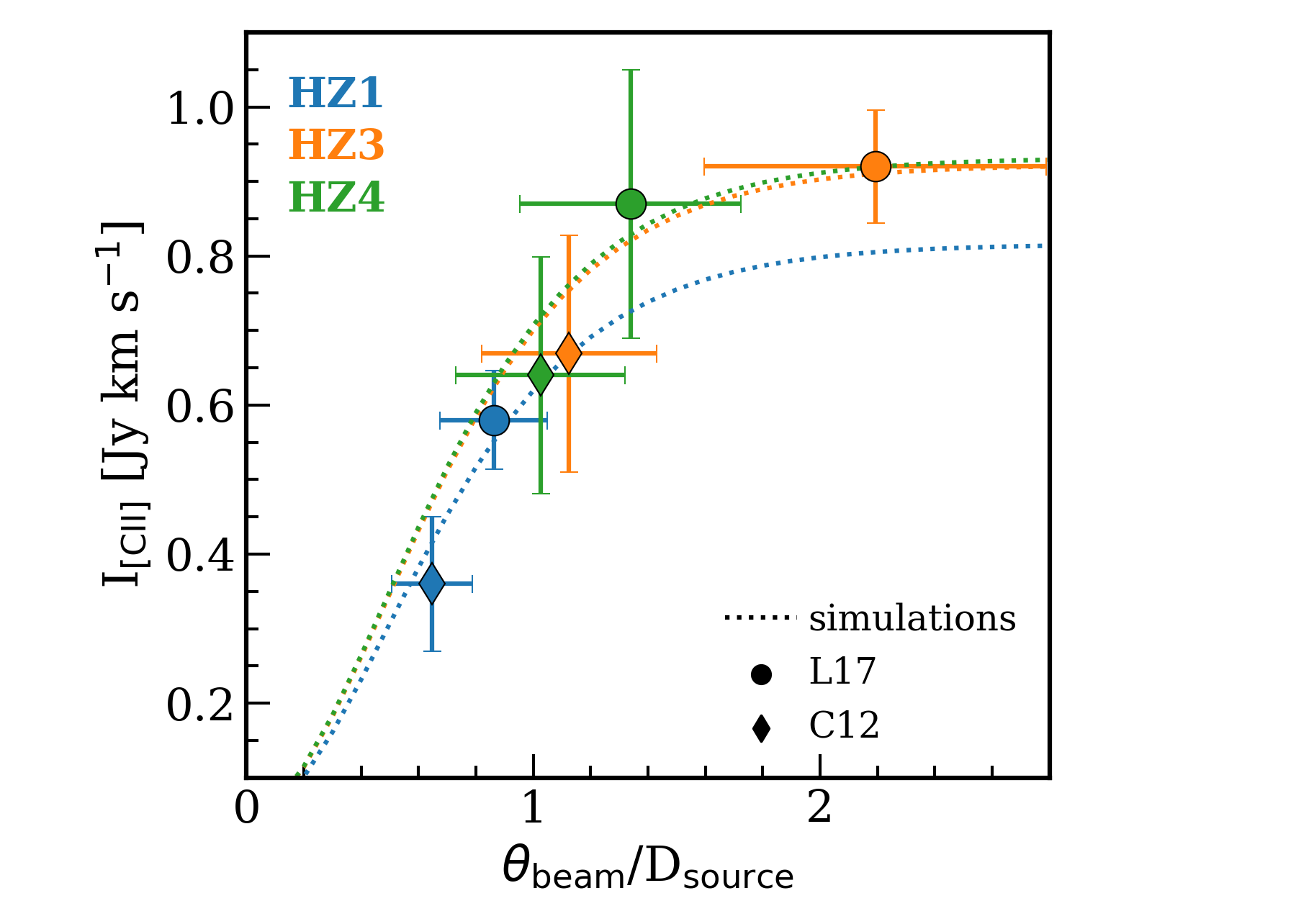}
    \caption{Integrated \cii\ flux (I$_{\rm [CII]}$) vs. ALMA angular resolution ($\theta_{\rm beam}$) normalised to the source size (D$_{\rm source}$). The measurements of HZ1, HZ3, and HZ4 are indicated with blue, orange, and green marks, respectively. The circles  show the estimates obtained from the 2017.1.00428.L dataset (L17; \citealt[][]{Le-Fevre:2019}), while diamond marks indicate the values from the 2012.1.00523.S program (C12; \citealt[][]{Capak:2015}). The predictions from our simulations are represented by the dotted lines. }
    \label{fig:sim_vs_obs}
\end{figure}

 Fig.~\ref{fig:sim_vs_obs} shows the \cii\ fluxes of HZ1, HZ3, and HZ4 inferred from the two individual datasets as a function of the $\theta_{\rm beam}$ normalised by the extent of \cii\ emission (D$_{\rm source}$)  estimated from the observations (see  App.~\ref{sec:app_hz} for more details).
 We notice  that the \cii\ measurements obtained from L17,  which has lower angular resolution, are systematically higher than those estimated in the higher angular-resolution images of C12. 
 This discrepancy is due to the two different ALMA beams.
Indeed in the  C12 program the angular resolution of the observations is comparable to the \cii\ size, so increasing the effect of SBD on the flux measurement. Although the significance of the \cii\ detection in C12  is \snr~$>5$, the sensitivity is not sufficiently high to recover the  diffuse and extended emission, whereas in L17 dataset the \cii\ line is barely resolved resulting in a higher flux estimate.

In Fig.~\ref{fig:sim_vs_obs}, we also report the predictions from our simulations. For each target we normalise the prediction curve to the weighted mean flux at the average $\theta_{\rm beam}$/D$_{\rm source}$ values of the two datasets.  The simulations are fully consistent with the observations, predicting that we miss 20-40\% of the total flux when the ALMA beam size is similar to the \cii\ extension ($\theta_{\rm beam}$/D$_{\rm source}\approx1$) and the \snr\ is relative low ($<10$).
We further note that total flux emission is fully recovered when the ALMA beam is two times larger than the source size.
The comparison of the observations and mock data shows that the predictions from our simulations can be used to derive the intrinsic flux of the data when the resolution and sensitivity are not sufficient to retrieve the total emission. 
In those galaxies where both \cii\ and rest-frame UV reveal a multi-component morphology, the correction factor can be applied to each individual component with its relative size. The main limitation of our predictions is that the correction factors suffer from large uncertainty and the total flux cannot be estimated with an error lower than $\sim20\%-30\%$. 
Moreover, our correction factors may not be appropriate for very complicated morphologies. In those cases,  ALMA simulations of the target provide a valid alternative to predict the missing flux due to the surface brightness dimming effect.

\section{Discussion}
\label{sec:discussion}

%\subsection{Extent of \cii\ emission}

In Sec.~\ref{sec:alma_data} we find that in our $z=6-9$ sample the extent of \cii\ is systematically larger than the \oiii\ line. Our findings parallel earlier $z>4$ galaxy morphological results revealing that \cii\ arises from an area typically 2-3 times more extended than the UV-emitting region. In some cases the \cii\ emission appear to be  extended even up to 6-10 kpc, which corresponds to 1\arcsec-1.3\arcsec\ at  $z\sim6$ \citep{Carniani:2018, Matthee:2017, Matthee:2019, Fujimoto:2020}.  

The origin of a such extended \cii\ structure is still debated. The diffuse emission can be ascribed to: (a) circumgalactic gas which is illuminated by the strong radiation field produced by the galaxy \citep{Carniani:2017, Carniani:2018, Fujimoto:2020}; (b) satellites in the process of accreting \citep{Pallottini:2017,Carniani:2018a, Carniani:2018,Matthee:2019}; (c) outflow remnants, which enriched the circum-galactic medium \citep{Maiolino:2015, Vallini:2015, Gallerani:2018, Fujimoto:2019, Fujimoto:2020, Pizzati:2020, Ginolfi:2020}. 
Despite its debated origin, it is clear that \cii\ is tracing gas on galactic scales different from those of rest-frame UV and \oiii\ emission. These different sizes should be taken into account in our measurements.

To date, most of the $z>6$ ALMA \cii\ observations have been carried out with semi-compact array configurations, leading to $0.5\arcsec<\theta_{\rm beam}<1.0\arcsec$. Such angular resolutions are sufficient to spatially resolve \cii\ emission, and thus reduce the surface brightness within the ALMA beam. Due to this surface brightness dimming effect, 
ALMA programs might have missed a fraction of diffuse emission resulting into a low \cii\ total luminosity.
In the following we discuss the effect of the SBD on the \lcii-SFR relation and \cii/\oiii\ luminosity ratio estimates.

% it has been unexpectedly found that the \cii\ is in general 2-3 times more extended than the UV regions \citep{Carniani:2018, Fujimoto:2020} and, therefore, its emission could be spatially resolved in ALMA observations in which extended or semi-compact array configurations ($\theta_{\rm beam}<0.8\arcsec-1\arcsec$)  have been adopted. This is also supported by \cite{Fujimoto:2019} and \cite{Ginolfi:2020}, which used stacking of ALMA data to show that the radial profile of the \cii\ line is extended out to about 10 kpc 
%Based on current morphological analysis of the FIR lines detected in $z>6$ star-forming galaxy, it is thus fundamental to quantify the effect of  angular resolution on FIR continuum and line luminosity measurements in order to investigate \cii-SFR relation and the \oiii/\cii relation in the EoR. 
%In the previous Sec. we showed that the lack of a strong \cii\ emission could be due to SBD effect related to the extended \cii\ emission. This effect could lead to a  \cii\ deficit placing the galaxy below the local \lcii-SFR relation and increasing the   \cii/\oiii\ luminosity ratio.

%

\subsection{\cii\ as tracer of SFR in the EOR?}
\label{sec:relation}

Here we  investigate the \lcii-SFR relation at $z>6$ by taking into account the fraction of ``missing'' \cii\ emission on the \lcii\ estimate. In addition to the sample of galaxies discussed in Sec.~3, in this analysis  we  include all star-forming galaxies at $z>6$ observed with ALMA so far. %to account the  the We have thus investigated the 
%taken into account the effect of SBD on the  \cii\ luminosity estimate of entire $z\gtrsim6$ sample of \cii\ emitters, including the nine targets re-analysed in this work. 
%

\begin{figure}
	\includegraphics[width=\columnwidth]{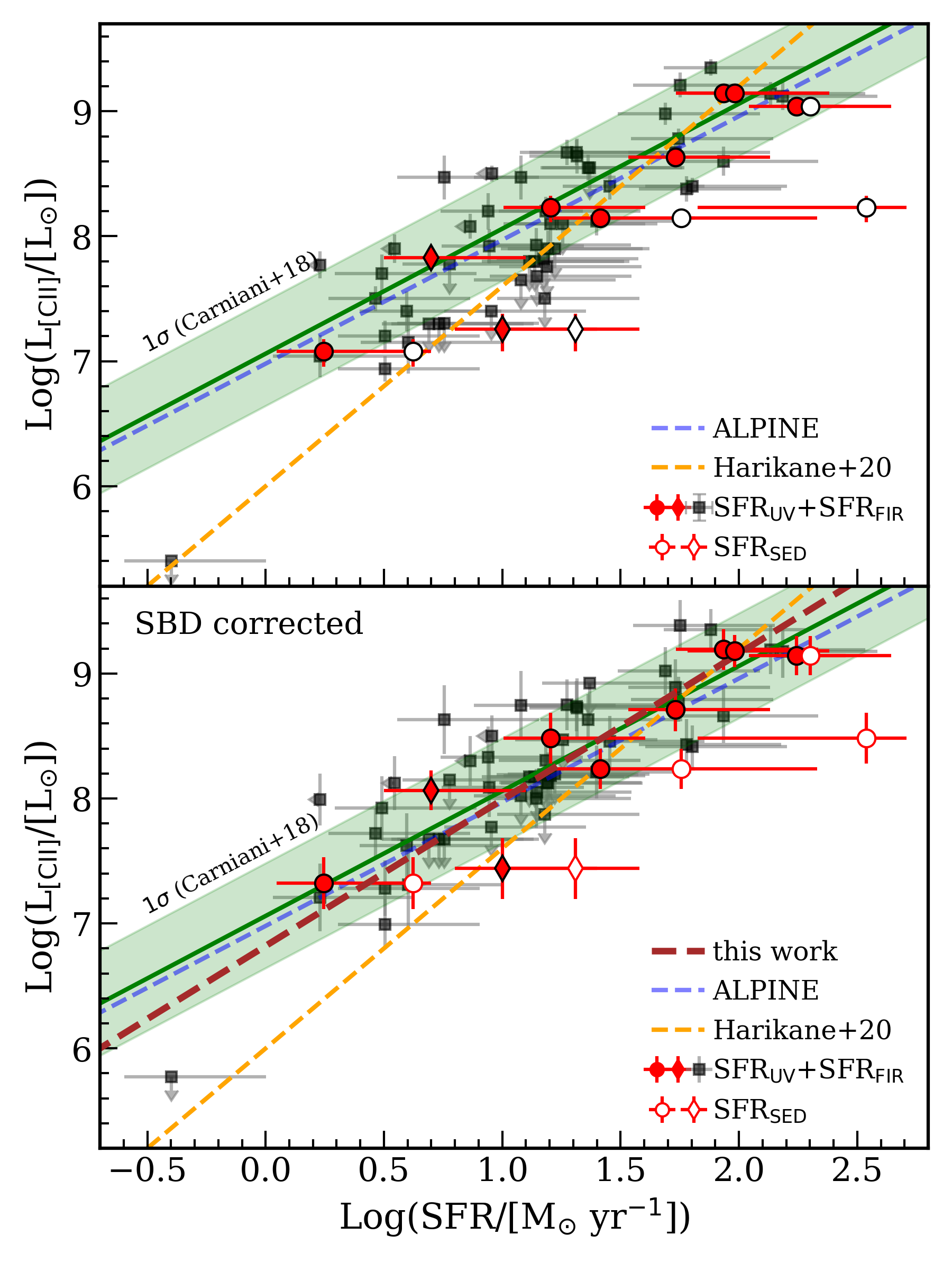}
    \caption{Top and bottom panels show the location of high-$z$ galaxies in the  \lcii-SFR before and after applying the correction for the SBD effect, respectively.
    Gray points show the location of all $z\gtrsim6$ star-forming galaxies reported in the literature so far, including lensed galaxies. Red marks represent the nine galaxies analysed in this work. The empty and filled symbols evidence the difference between the two SFR calibrations based on SED fitting and L$_{\rm UV}$+L$_{\rm FIR}$, respectively, whereas the diamonds distinguish the sources in which the \cii\ is not co-spatial with  UV emission. The green line shows the local relation by \citet{De-Looze:2014} for HII-like galaxies ($\log SFR = (1.00\pm0.04)\log L_{\rm [CII]} -(7.06\pm0.33)$), with the shaded area corresponding to the $1\sigma$ uncertainty for high-$z$ galaxies from \citet{Carniani:2018}. The shaded blue line represents the best-fit results from ALPINE survey by \citet{Schaerer:2020} ($\log L_{\rm [CII]} = (0.99\pm0.09)\log SFR+(6.98\pm0.16)$), while the orange line shows the result from $z>6$ galaxies by \citet{Harikane:2019}($\log L_{\rm [CII]} = 1.6\log SFR+6.0$). The dashed dark red line in bottom panel illustrates the best-fit results for the SBD-corrected data.} 
    \label{fig:lcii_sfr}
\end{figure}

Based on our simulations, we have corrected the observed \cii\ luminosities depending on the ALMA beam, SNR of the detection, and extent of \cii\ emission.
For the upper limits, where the extent of the  carbon emission is not known, we assume that the  \cii\ line is about two times larger than the UV \citep[][]{Carniani:2018, Fujimoto:2020}.
We have also derived the total SFR of each source in an uniform way. For those galaxies revealing continuum emission in the ALMA bands, the total SFR has been estimated by adding the SFR$_{\rm IR}$ based on $L_{\rm FIR}$ to the SFR$_{\rm UV}$ calculated from the UV luminosity. We have  used a modified blackbody with dust temperature $T_{\rm dust} = 40$~K and emissivity index $\beta=1.5$ to reproduce the FIR emission\footnote{As the $L_{\rm FIR}$ depends strongly on the assumed T$_{\rm dust}$ and $\beta$, we have associated an asymmetric uncertainty of -0.2 dex and +0.4 dex to the  $L_{\rm FIR}$ estimates, and thus to the SFR$_{\rm IR}$ and total SFR.} of all galaxies. On the other hand, we have assumed a total SFR$\approx$SFR$_{\rm UV}$ with no dust correction for those galaxies without continuum detection. Finally, we have taken into account the multi-components morphology of high-$z$ galaxies and performed  the proper associations between \cii\ and UV components.
Top and bottom panels of Fig.~\ref{fig:lcii_sfr} show the \lcii-SFR diagram  before and after applying the correction for the ``missing''  extended \cii\ emission.

After correcting for the SBD effect, the $z>6$ galaxies become more consistent with the local relation. Most of  the upper limits are within  the intrinsic dispersion of the relation. Interestingly, the lensed galaxy MS0451-H \citep{Knudsen:2016}, which is the only galaxy in the sample with  SFR~$<1$~\sfr, still appears to deviate from the local relation by more than 2$\sigma$. However, we notice that this source is a lensed arc with an UV extension of $\sim5-6$\arcsec, while the ALMA beam is only 1.6\arcsec, hence the flux could be spatially fully resolved (see App.~\ref{app:appb}). Deeper ALMA images of this source are therefore fundamental to recover the total emission and verify if the deviation from the local relation is real or not.

The resulting best fit for the SBD-corrected data is $\log L_{\rm [CII]} = (1.1\pm0.2)\log SFR + (6.8\pm0.2)$ that is consistent within the errors with both the local relation by \cite{De-Looze:2014} and the $z=4-5$ fitting result reported by \cite{Schaerer:2020}. We note that our results are in contrast with those shown in the  previous study by \cite{Harikane:2019} who find a steeper  \lcii-SFR relation at $z>6$ (orange curve in Fig.~\ref{fig:lcii_sfr}). The discrepancy between the two results mainly depends on the handling of \cii\ data. Indeed the SBD correction returns more conservative upper limits for \lcii\ and moves most of the \cii\ non-detections closer to the local relation. In addition to that, in our sample we have three new \cii\ detections that have been considered as non-detections in previous works. Finally, in our study we have  uniformly estimate the total SFR of each source based on the UV and FIR luminosity. These three effects explain the difference in normalisation and slope between the best-fit relation of our study and that of \cite{Harikane:2019}. New \cii\ observations of galaxies with SFR$<1-5$~\sfr\ will be fundamental to put stronger constrain on the slope of \lcii-SFR relation at $z>6$.

For the nine targets re-analysed in this work we show their location on the SFR-\lcii\ diagram by adopting both the SFR estimated from the UV+FIR luminosity, or from SED fitting, when available in the literature. We note that, if we use the SFR estimated from the UV+FIR luminosity \citep{Kennicutt:2012}, which is the same method used for the other \cii\ emitters from the literature \citep[e.g.][]{Matthee:2019} and for the ALPINE survey \citep{Schaerer:2020}, our sources are in agreement with the local relation, within the uncertainty of 0.48~dex defined by  \cite{Carniani:2018}.
On the other hand, if we adopt the SFR from SED fitting, high-$z$ galaxies appear systematically below the local relation. 
It is worth mentioning that the SED fitting method returns different SFR estimates with respect to those obtained from UV+FIR  calibrators because   the assumed  star-formation histories, dust-attenuation curves, and stellar population ages are different between the two methods \citep[e.g.][]{Schaerer:2013, Schaerer:2020, Faisst:2020}.
In this context, future observations in the near- and mid-IR with JWST \citep{Gardner:2006,Williams:2018,Chevallard:2019}, and SPICA \citep{Spinoglio:2017, Egami:2018}  will be crucial to better constrain the SED shape and, thus, determine galaxy properties as SFR. In the rest of the work, we use the SFR from  UV+FIR luminosity  since it is the same method used to determine the local \lcii-SFR relation \citep{De-Looze:2014}.

\begin{figure}
	\includegraphics[width=\columnwidth]{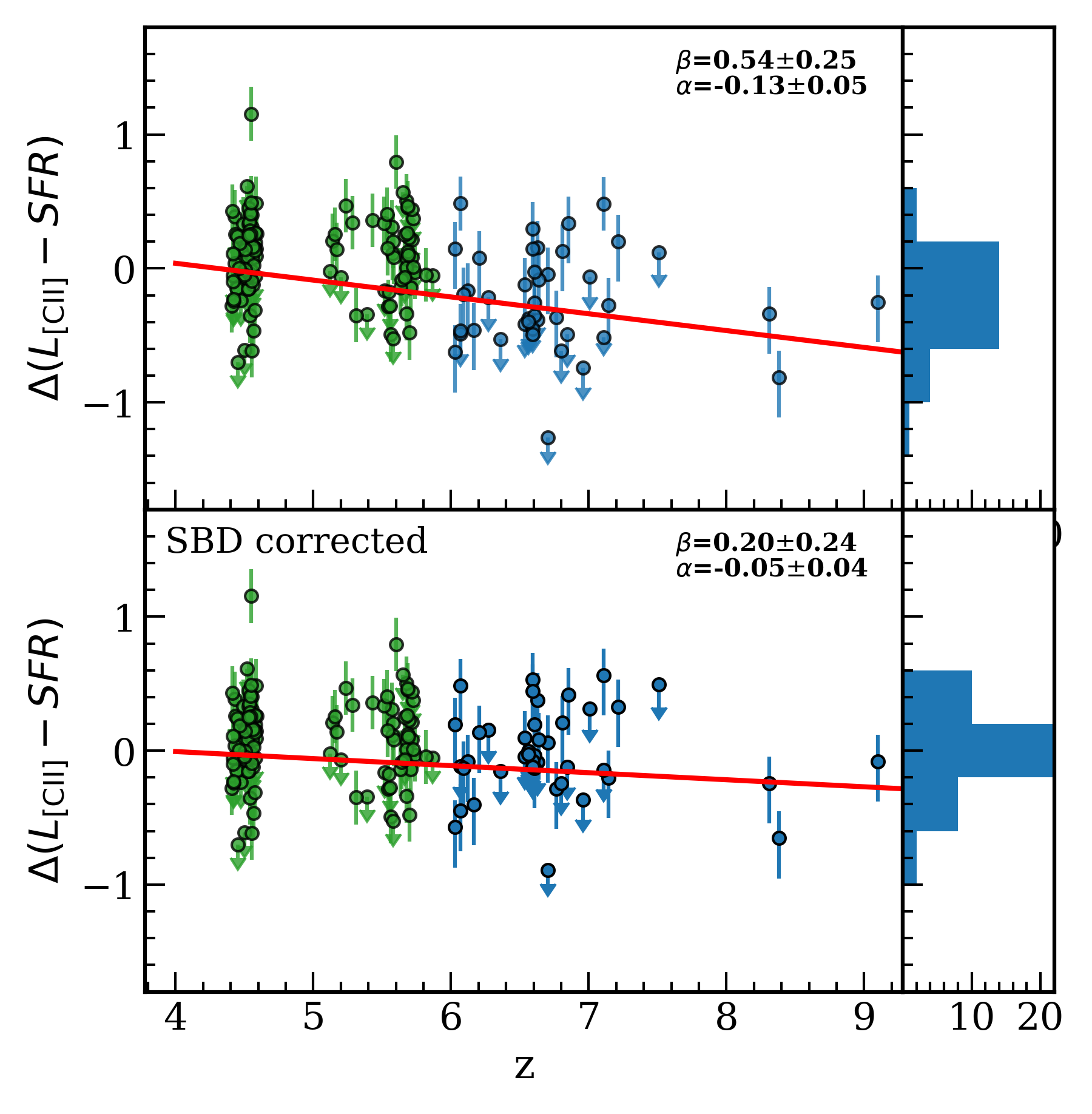}
    \caption{Top and bottom panels show the deviation from the local \lcii-SFR relation as a function of redshift  before and after applying the correction for the SBD effect, respectively. Blue circles represents $z\gtrsim6$ star-forming galaxies, while green circles are ALPINE galaxies ($4<z<5.5$; \citealt{Bethermin:2020}).
    Red lines indicate our linear fit and the best-fit parameters are reported in the top-right corner.  The top an bottom  histograms represent the distributions from $z\gtrsim6$ sample  before and after applying the correction for the SBD effect, respectively}
    \label{fig:deviation}
\end{figure}

In Fig.~\ref{fig:deviation} we report the offsets from the local \lcii-SFR relation for the $z\gtrsim6$ galaxies (blue circles) and ALPINE sample (green marks)  as a function of redshift before and after correcting for the SBD effect.  
A linear fit of the total sample gives: 
\begin{equation}
\Delta = (-0.13\pm0.05)z-(0.54\pm0.25) \,
\end{equation}
for the uncorrected galaxies (top panel) and 
\begin{equation}
\Delta = (-0.05\pm0.04)z-(0.20\pm0.24) \,
\end{equation}
for the corrected sample (bottom panel).
The large uncertainties on the best-fit values suggest that there is a no or very weak correlation with the redshift, as also observed at $4<z<5.5$ by the ALPINE survey \citep{Schaerer:2020}. However the low statistics at $z>7$ does not allow us  to determine definitively whether the \lcii-SFR relation evolves with redshift or not.

In conclusion, if we correct for the ``missing'' \cii\ emission and estimate the SFR from  UV+FIR luminosity, the whole $z\gtrsim6$ sample but one seems to be in agreement with the local relation.
More specifically the average offset from the local relation  is $\Delta^{z=6-9} = -0.07\pm0.3$ for the corrected sample and $\Delta^{z=6-9} = -0.2\pm0.3$ for the uncorrected one.
The small offset indicates that the $z>6$ targets observed with ALMA so far  are not extremely metal-poor galaxies ($Z<0.2Z_\odot$); otherwise, we would expect to observe a clear deviation from  \lcii-SFR relation as indicated by models and simulations \citep{Vallini:2015, Pallottini:2017, Pallottini:2019, Ferrara:2019, Lupi:2020}.

It is worth stressing that the intrinsic dispersion of the \lcii-SFR relation observed at $z>4$ is 0.42-0.48~dex \citep{Carniani:2018,Schaerer:2020}, two times larger than that inferred from the local HII-like  star-forming galaxies  \citep[0.28~dex][]{De-Looze:2014}.  
Such broad dispersion is indicative of a broader range of ISM properties spanned by such distant galaxies with respect to the local population. 
In this context, the spatially resolved  $\Sigma_{\rm SFR}-\Sigma_{\rm [CII]}$ provides a better comparison between high-$z$  and local galaxies, since it is more sensitive to the ISM properties  \citep{Ferrara:2019}. The \cii\ surface brightness at $z>6$ is  systematically lower  than that expected  from nearby galaxies \citep[see Fig.~10 by][]{Carniani:2018}. This deficit, which is not visible in the integrate \lcii-SFR, may indicate that $z>6$ galaxies deviate from the Kennicutt-Schmidt relation \citep{Pallottini:2019,Ferrara:2019}, as recently confirmed in \cite{Vallini:2020} for a galaxy at $z=6.8$ by combining  the observations of the \cii\ and the rest-frame UV line of C$\scriptstyle\rm III$]. Future ALMA surveys should therefore aim at investigating the  spatially resolved relation  by performing very deep and high-angular resolution observations.

\subsection{\oiii\ over \cii\ luminosity ratio}
\label{sec:line_ratio}
\begin{figure}
	\includegraphics[width=\columnwidth]{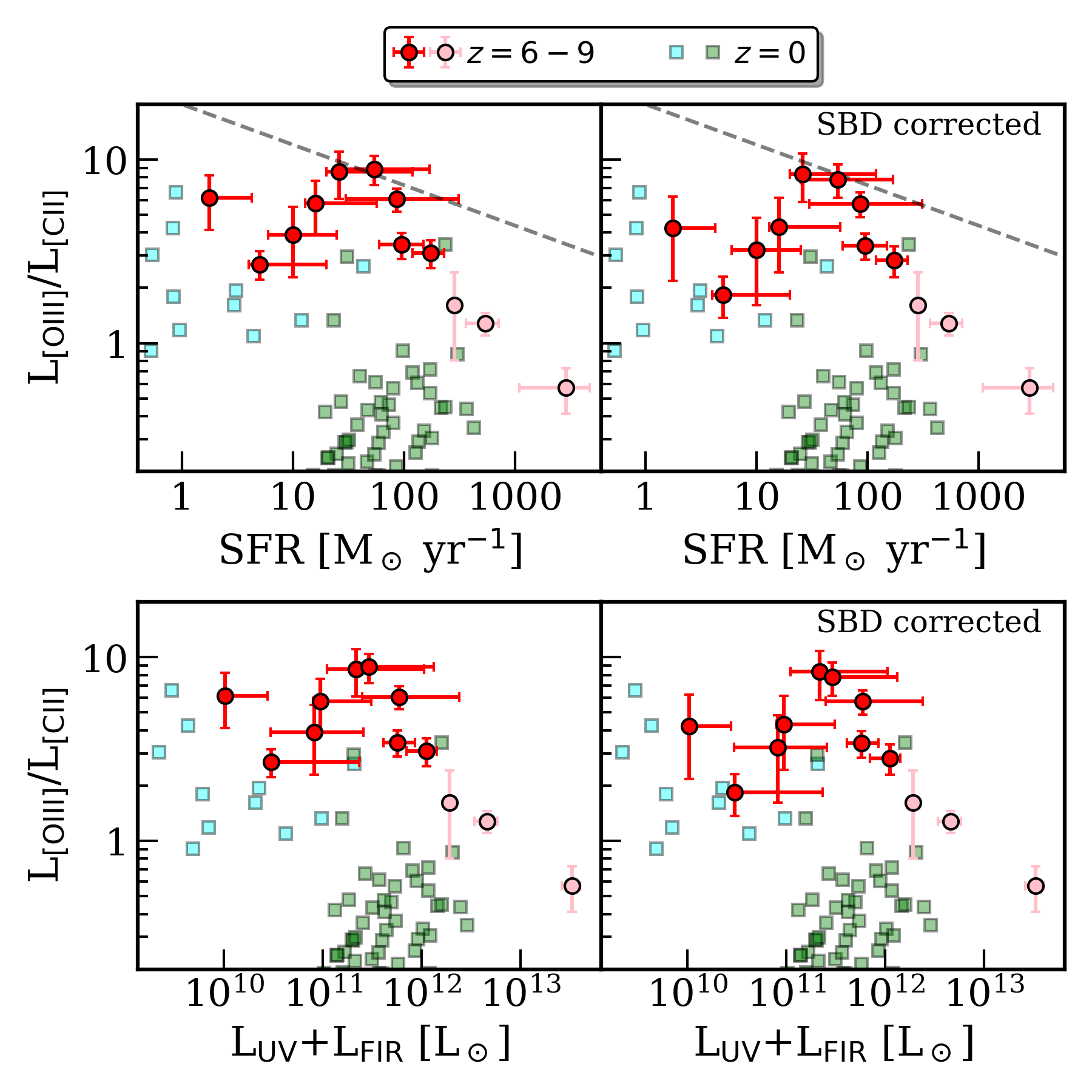}
    \caption{\loiii/\lcii\ ratio as a function of SFR (top) and L$_{\rm UV}$+L$_{\rm FIR}$ (bottom). In the left and right panels, we show the luminosity line ratio before and after applying the correction for the SBD effect, respectively. Measurements for the nine $z>6$ star-forming galaxies analysed in this work are shown with red circles. Literature results for $z>6$ dusty star-forming galaxies (\citealt{Marrone:2018, Walter:2018}) are reported as pink circles. Blue and green squares show the typical ratios observed in local metal-poor and metal-rich star-forming galaxies, respectively (\citealt{De-Looze:2014, Cormier:2015, Diaz-Santos:2017}). The grey line represents the best-fitting results for the $z>6$ galaxies by \citet{Harikane:2019}.} 
        \label{fig:oiii_cii_ratio}
\end{figure}

Fig.~\ref{fig:oiii_cii_ratio} shows the $L_{\rm [OIII]}/L_{\rm [CII]}$ ratio as a function of  SFR (top panels), and  bolometric luminosity defined as L$_{\rm UV}$+L$_{\rm FIR}$ (bottom panels). In the left panels, the line ratios have been estimated directly from the measurements (col.~5 of Table~\ref{tab:summary}), while in the right panels the line luminosities have been corrected for the ``missing'' extended emission due to the effect of the SBD  (col.~6 of Table~\ref{tab:summary}). The correction factors have been determined from our simulations Table~\ref{tab:corr}), depending on the angular resolution, emission size, and \snr\ of the detections. 

All high-$z$ star-forming galaxies detected both in \oiii\ and \cii\ exhibit luminosity ratios $> 2$ , i.e. higher than the average value reported for local metal-rich star-forming galaxies \citep{De-Looze:2014, Cormier:2015, Diaz-Santos:2017}.
However, differently from previous studies, we do not find any value higher than 10. In particular, after the correction for the SBD effect, the luminosity ratios span a range between 1 and 8, that is more consistent with local dwarf galaxies \citep[$1<$~\loiii/\lcii~$< 10$;][]{Madden:2013,Cormier:2015} and simulations \citep[$0.5<$~\loiii/\cii~$< 10$;][]{Pallottini:2017a, Pallottini:2019, Katz:2019, Arata:2020, Lupi:2020b}

Fig.~\ref{fig:oiii_cii_ratio} also points out the large scatter present in our sample, with no clear dependence of the luminosity ratio on either SFR or L$_{\rm bol}$($=L_{\rm UV}+ L_{\rm IR}$),
in contrast to  earlier studies \citep[e.g.][]{Hashimoto:2019,Harikane:2019, Bakx:2020}.
Indeed the apparent decreasing trend in \loiii/\lcii\ for increasing SFR (and  L$_{\rm bol}$) proposed in previous works (gray dashed line in top panels) was probably driven by extreme ($>10$) line ratios quoted in some galaxies with low SFR. However, our re-analysis of ALMA data shows that faint galaxies have a \loiii/\lcii\ similar to the line ratios observed in the whole sample. 
In addition to that, it is worth mentioning that  in Fig.~\ref{fig:oiii_cii_ratio} we adopt a uniform approach to estimate the total SFR of each galaxy, which is based on the L$_{\rm UV}$ and  L$_{\rm IR}$, while previous work by \cite{Harikane:2019} reports the SFR estimates directly from literature.

As discussed in \cite{Harikane:2019}, the \loiii/\lcii\ depends mainly on ISM properties such as PDR covering fraction, density, metallicity, and C/O abundance ratio. Therefore, a weak (or absent) relation with the total SFR is not surprising. 
However, we expect a dependence with the intensity radiation field as the line ratio intensity should increase with the ionisation parameters \citep{Harikane:2019}. 
Indeed \citet{Pallottini:2019} show the \loiii/\lcii\ ratio reaches values as high as 10 in the central region of their simulated galaxies, where the  intensity radiation field  has maximum intensity. In the light of these results future deep and high angular-resolution ALMA observations will be fundamental to investigate the spatially resolved \loiii/\lcii\ ratio in high-$z$ galaxies and compare it with the  SFR surface density map. 

By using  Cloudy calculations, \cite{Harikane:2019} find that the \loiii/SFR and \lcii/SFR ratios obeserved in $z>6$ can be reproduced with PDR covering fractions of 0-10\% and ionisation parameters  $-2\lesssim\log U_{\rm ion}\lesssim-1$, which is 10-100 times higher than what observed in local galaxies. Our new \cii\ detections, SBD correction and uniform total SFR calculations for the $z>6$ galaxies do not change previous results. The new \lcii/SFR estimates locate the high-$z$ galaxies in the range $6.4< \log(L_{\rm [CII]}/SFR)<7.4$, that is slightly higher than that reported by \cite{Harikane:2019}, but $\log U_{\rm ion}\approx -2$ and a PDR covering  fraction of 10\%  are still requested to reproduce \lcii/SFR ratios. As discussed in \cite{Harikane:2019} such a low PDR covering fraction  implies the presence of escape routes for the Ly$\alpha$ photons and most importantly for the Lyman continuum photons that are responsible of reionization of the Universe. In conclusion such galaxies can play an important role in the reionization process.

\section{Summary and Conclusions}
\label{sec:conclusion}

In the last few years, several ALMA programs  have reported weak \cii\ emission (or upper limits) for   star-forming galaxies at $z>6$. These low luminosities have suggested that the \lcii-SFR relation distant Universe might be different from the local one. Recent  ALMA \oiii\ observations have also revealed that the \loiii/\lcii\ line ratio at $z>6$ is systematically higher than that observed in low-$z$ metal-poor and metal-rich galaxies.

Here we  have investigated if the surface brightness dimming (SBD), caused by the combination of the spatially extension of the \cii\ emission  and the relative high-angular resolution of current ALMA observations, could be the origin of the \cii\ deficit reported in earlier works. The main results of our analysis are summarised below.
\begin{itemize}

\item We have analysed the ALMA observations of the  nine  $z=6-9$ targets observed both in \cii\ and \oiii. 
By performing different \emph{uv}-tapering to optimise the detection of the diffuse emission, we have identified the \cii\ line in the whole sample with a level of significance of $\simgt 4\sigma$ at the location and redshift of the either \oiii\ or \lya\ line.
We have found that \textit{{\rm [CII]} emission is systematically $\simeq 2\times$ more extended than {\rm [OIII]} one}. This result is in line with other works showing the effective radius  of \cii\ is in  general larger than the radius of UV region by a factor 2-3 \citep{Carniani:2018, Fujimoto:2020}. The  origin of such extended component is yet unknown; an appealing explanation is that it may be associated to either inflowing or outflowing material \citep{Maiolino:2015, Carniani:2017, Carniani:2018, Gallerani:2018, Pizzati:2020, Fujimoto:2020, Ginolfi:2020}.

\item The extended \cii\, emission might lead to a significant underestimation of the total FIR line luminosity in spatially resolved ALMA observations. By performing ALMA simulations with different array configurations and exposure times, we conclude that \textit{$20-40$\% of the total} \cii\ \textit{flux might be missed when the angular resolution is comparable to the size of the emitting region}, even when the level of significance of the line is $\sim 5-10 \sigma$.  The fraction of missing flux increases up to 70\% when the line is detected with a \snr$<5$.

\item Using our simulations, further validated against observations, we compute the missing \cii\ emission both in our sample and in all galaxies so far observed by ALMA at $z>6$. We have thus investigated the offset from the local \lcii-SFR relation.
On average, $z>6$ galaxies with SFR~$>5~{\rm M_\odot~yr^{-1}}$ are located slightly below the local relation ($\Delta^{z=6-9}=-0.07\pm0.3$), but within the intrinsic dispersion of the relation at high-$z$. This agrees well with results at $4<z<6$ by \cite{Schaerer:2020}, suggesting  little evolution of the \lcii-SFR relation with redshift.
However we also note that the  low  statistics  at $z>7-8$  does  not  allow us to determine definitively whether the  relation evolves at very high-$z$ or not.

\item  We notice that if we adopt the SFR from SED fitting rather than that based on UV+IR luminosity, $z>6$ star-forming galaxies are systematically offset from the local relation. 
Therefore future JWST and SPICA observations will be fundamental to put constrains on the SFR, and determine if these galaxies are indeed  \cii\ faint.

\item The new \cii\ detections exclude \oiii/\cii\ luminosity ratios $>10$ and, once \cii\ is corrected for SBD, we find  $2<$~\loiii/\lcii~$<8$, in much better agreement with local dwarf galaxies and simulations. Differently from previous works, we do not confirm a dependence of  \loiii/\lcii\ on SFR and bolometric luminosity, i.e. L$_{\rm UV}$+L$_{\rm FIR}$. As shown by \cite{Harikane:2019}, we suggest that that \loiii/\lcii\ is more related to the local properties of the ISM (e.g. gas metallicity, density, PDR covering fraction)  rather than to global galaxy properties.
\end{itemize}

In summary, the SBD caused by the spatially resolved ALMA observations could have a strong impact on our flux line measurements, and lead to spurious non-detections. 
The {\it uv}-tapering is a possible alternative to recover the "missing" extended emission to the detriment of sensitivity. 
However, this method is not sufficient to infer the total emission arising from the galaxy.
Moreover, the missing  \cii\ emission problem due to the high resolution (and low sensitivity) observations could have an even larger impact in lensed galaxy arcs, since their extension on the sky could be several times the ALMA beam due to gravitational magnification. Lensed galaxies with $\mu>5-10$ may go undetected because the \cii\ flux is resolved out, and its surface brightness drops below the  detection limit. 

Future ALMA programs at low resolutions ($>1\arcsec$)  will be ideal to (i) recover the extended \cii\  emission component, (ii) investigate both the \lcii-SFR relation, and (iii) \oiii/\cii\ luminosity ratio.
On the other hand, high-angular resolution ALMA observations with high sensitivity are crucial to study the excitation of the FIR line emission in early galaxy systems, and accurately map the ISM properties in spatially resolved regions.

\section*{DATA AVAILABILITY}
The data used in the paper are available in the ALMA archive at https://almascience.nrao.edu.
The derived data and simulations generated in this research will be shared on reasonable request to the corresponding author.

\section*{Acknowledgements}
We thank the anonymous referee for constructive comments and suggestions.
This paper makes use of the following ALMA data:\\
 ADS/JAO.ALMA\#2012.1.00523.S, ADS/JAO.ALMA\#2012.A.00040.S, ADS/JAO.ALMA\#2012.A.00374.S,
ADS/JAO.ALMA\#2013.A.00021.S
ADS/JAO.ALMA\#2016.1.00856.S,
ADS/JAO.ALMA\#2017.1.00428.L,
ADS/JAO.ALMA\#2017.1.00697.S,
ADS/JAO.ALMA\#2017.A.00026.S.
 ALMA is a partnership of ESO (representing its member states), NSF (USA) and NINS (Japan), together with NRC (Canada), MOST and ASIAA (Taiwan), and KASI (Republic of Korea), in cooperation with the Republic of Chile. The Joint ALMA Observatory is operated by ESO, AUI/NRAO and NAOJ.
AF, SC, and AL acknowledge support from the ERC Advanced Grant INTERSTELLAR H2020/740120. 
RM acknowledges support from the ERC Advanced Grant 695671 `QUENCH' and from the Science and Technology Facilities Council (STFC). This work reflects only the authors' view and  the  European Research Commission is not responsible for information it contains. EV acknowledges INAF funding from the program ``Interventi aggiuntivi a sostegno della ricerca di main-stream (1.05.01.86.31)".
%
%
%%%%%%%%%%%%%%%%%%%%%%%%%%%%%%%%%%%%%%%%%%%%%%%%%%


\bibliographystyle{mnras}
\bibliography{biblio_oiii_cii.bib} % if your bibtex file is called example.bib

%%%%%%%%%%%%%%%%%%%%%%%%%%%%%%%%%%%%%%%%%%%%%%%%%%

%%%%%%%%%%%%%%%%% APPENDICES %%%%%%%%%%%%%%%%%%%%%

\appendix

\section{ALMA observations}

In Table~\ref{tab:appendix} we list the properties of ALMA observations and FIR line detections analysed in this work. For some galaxies we report the values from previous works, since our re-analysis returns the same results within the errors.% We thus encourage users of these estimates to cite the primary sources as well.

\begin{landscape}
 \begin{table}
  \caption{\cii\ and \oiii\ line properties}
  \label{tab:appendix}
  \begin{tabular}{lccccccccc}
    \hline
     &  MACS1149-JD1 &  A2744-YD4 &  MACSJ0416-Y1 &  SXDF-NB1006-2 &  B14-65666 &  BDF-3329 &  J0217  & J0235 &  J1211 \\
    
    \hline
    & \multicolumn{9}{c}{\cii}\\
    \\
     $\theta_{\rm beam}$  & $1.4\arcsec\times1.3\arcsec$ & $1.5\arcsec\times0.9\arcsec$ & $0.64\arcsec\times0.46\arcsec$ & $0.6\arcsec\times0.5\arcsec$ & $0.29\arcsec\times0.23\arcsec$ &  $0.6\arcsec\times0.5\arcsec$ & $0.7\arcsec\times0.7\arcsec$ & $0.8\arcsec\times0.7\arcsec$ & $0.8\arcsec\times0.6\arcsec$  \\

    {\it uv}-taper & 1.0\arcsec & 0.2\arcsec & 0.27\arcsec &  0.4\arcsec & - & - & - & - & -\\
    SNR$_{\rm peak}$ & 3.8 & 4.7 & 6.5 & 4.1 &  7 & 5.2 & 10 & 8.0 & 11\\
    $z_{\rm [CII]}$ & $9.1099\pm0.0016$ & $8.3796\pm0.0002$ & $8.31132\pm0.00037$ & $7.2127\pm0.0009$ & $7.1521\pm0.0004$ & $7.109\pm0.001$  & $6.2033 \pm 0.0009$ & $6.0894\pm0.0010$ & $6.0291\pm0.0008$\\
    FWHM [\kms] & $130\pm110$ & $50\pm16$ & $191\pm29$ & $230\pm80$  & $349\pm31$ & $100\pm30$ & $316\pm117$ & $270\pm135$ & $170\pm98$\\
    S$\Delta$v [mJy \kms] & $66\pm18$ & $22\pm5$ & $120.2\pm20.4$ & $130\pm30$  & $870\pm11$ & $52\pm7$ & $1.36\pm0.20$  & $0.42\pm0.07$ & $1.42\pm0.15$ \\
    extent$^{\dagger}$ & $1.8\arcsec\pm1.1\arcsec$ & $1.6\arcsec\pm0.9\arcsec$ & $0.48\arcsec\pm0.14\arcsec$ & $1.6\arcsec\pm0.7\arcsec$ &  $0.85\arcsec\pm0.11\arcsec$ & $1.1\arcsec\pm0.8\arcsec$  & $1.35\arcsec\pm0.19\arcsec$ & $1.0\arcsec\pm0.3\arcsec$ & $1.35\arcsec\pm0.17\arcsec$\\
    \hline
    & \multicolumn{9}{c}{\oiii}\\
    \\
        $\theta_{\rm beam}$  & $0.62\arcsec\times0.52\arcsec$  & $0.23\arcsec\times0.17\arcsec$ & $0.26\arcsec\times0.21\arcsec$  & $0.39\arcsec\times0.37\arcsec$ & $0.35\arcsec\times0.36\arcsec$ &  $0.5\arcsec\times0.4\arcsec$ & $0.7\arcsec\times0.6\arcsec$ & $0.7\arcsec\times0.6\arcsec$ & $0.8\arcsec\times0.6\arcsec$  \\

    {\it uv}-taper & - & - & 0.35\arcsec &  - & - & - & - & - & -\\
    SNR$_{\rm peak}$ & 7.4 & 4.0 & 6.0 & 5.2 &  9 & 5 & 12 & 12 & 11\\
    $z$ & $9.1096\pm0.0006$ & $8.382\pm0.001$ & $8.3118\pm0.0003$ & $7.2120\pm0.0003$  & $7.1521 \pm 0.0004$ & $7.117\pm0.001$ & $6.2044\pm0.0013$ & $6.0906\pm0.0009$ & $6.0295\pm0.0009$\\
    FWHM [\kms] & $154\pm39$ & $49\pm4$ & $141\pm21$ & $80$  & $1.50\pm0.18$& $40\pm10$ & $194\pm123$ & $389\pm117$ & $374\pm162$\\
    S$\Delta$v [mJy \kms] & $49\pm12$ & $17\pm10$ & $660\pm160$ & $450\pm90$  &  $429\pm37$ & $85\pm12$ & $4.57\pm1.06$  & $2.10\pm0.18$ & $2.69\pm0.40$ \\
    extent$^{\dagger}$ & $0.8\arcsec\pm0.3\arcsec$ & $<0.2\arcsec$ & $0.3\arcsec\pm0.2\arcsec$ & $0.4\arcsec\pm0.2\arcsec$  & $0.66\arcsec\pm0.09\arcsec$ & $<0.4\arcsec$ & $0.74\arcsec\pm0.10\arcsec$ & $0.53\arcsec\pm0.15\arcsec$ & $0.74\arcsec\pm0.17\arcsec$\\
    \hline
  \end{tabular}
  
  {\bf Note}. ${\dagger}$ major axis FWHM size.
 \end{table}
\end{landscape}
%\end{verbatim}

\section{Blind-line search method}\label{app:blind}
The blind-line search method used in this work is based on a customised line finder code written in Python and optimised for ALMA observations. The code generates channel maps with different input line widths by averaging the ALMA datacube in frequency. In each channel map the algorithm estimates the noise level and searches for peaks exceeding a fixed SNR threshold, saving their properties (e.g. frequency, position) in a temporary file.  Finally the code performs a cross-match between all the extracted candidates to identify duplicates with similar position and frequency, and keeps only the candidates with the highest SNR.

For our study, we have searched candidate lines within 1000 \kms\ from the \oiii\  and within 5\arcsec\ from the UV counterpart.  As the \cii\ line width is not known a priori, we have used a set of different line widths from 50 \kms\ to 500 \kms. 
The SNR threshold of the peaks has been fixed to 3.6 that is the same limit used in the \cii\ ALPINE survey \citep{Bethermin:2020}. Finally, we have set a distance of one ALMA beam and $|\Delta v = 500|$~\kms\ to discriminate duplicates.  

We note that our line finder code  is optimised to detect point-source emission. To detect extended emission, we have generated  {\it uv}-tapered datacubes with different angular resolution for each target and run our customised code on them (Sec.~\ref{sec:obs}).

\section{Results from ALMA simulations}

Here we report  the results from the simulations with different exposure times and ALMA array configurations. Table~\ref{tab:corr} shows the ratio between the measured flux density and the input model as a function of  signal-to-noise ratio  of the detection and the angular resolution normalised by the source size.

\begin{table*}
    
    \caption{Ratio between measured  and intrinsic flux as a function of ALMA angular resolution normalised by the {\cii\ source size} ($\theta_{\rm beam}/{\rm D_{\rm source}}$) and signal-to-noise ratio (SNR)}
	\label{tab:corr}
	\centering

		\begin{tabular}{|c|c||c|c|c|c| c|}
		
		\hline
		\multicolumn{2}{|c||}{} & \multicolumn{4}{c|}{$\theta_{\rm beam}/{\rm D_{\rm source}}$} \\
		\multicolumn{2}{|c||}{} & 2.3 & 1.8 & 1.1 & 0.7 &  0.4 \\
		\hline
    	\multirow{6}{*}{\rotatebox[origin=c]{90}{\parbox[c]{1cm}{\centering SNR}}} &   3 & $0.68\pm0.24$ & $0.73\pm0.27$ & $0.63\pm0.15$ & $0.40\pm0.10$ & $0.28\pm0.08$ \\
        & 5 & $0.84\pm0.24$ & $0.86\pm0.22$ & $0.78\pm0.21$ & $0.64\pm0.18$ & $0.57\pm0.15$ \\
        &7 &$0.88\pm0.16$ &$0.85\pm0.14$ &$0.81\pm0.14$ & $0.80\pm0.14$ &$0.79\pm0.11$ \\
        &9 &$0.88\pm0.13$ &$0.90\pm0.14$ &$0.88\pm0.16$ &$0.87\pm0.12$ &$0.88\pm0.08$ \\
        &11 &$0.91\pm0.12$ &$0.90\pm0.12$ &$0.94\pm0.13$ &$0.92\pm0.09$ & $0.94\pm0.07$ \\
        &13 &$0.94\pm0.10$ &$0.96\pm0.10$ &$0.96\pm0.10$ &$0.95\pm0.08$ &$0.97\pm0.06$ \\ &15 &$1.01\pm0.10$ &$0.98\pm0.10$ &$0.99\pm0.08$ &$0.96\pm0.07$ &$1.0\pm0.06$ \\

    \end{tabular}
\end{table*}

\section{Data calibration and analysis of HZ1, HZ2, and HZ4}\label{sec:app_hz}

After retrieving the two ALMA datasets, 2012.1.00523.S \citep{Capak:2015} and 2017.1.00428.L \citep{Le-Fevre:2019, Bethermin:2020}, we have reduced the data by adopting the appropriate \verb'CASA' pipeline version and performed the \cii\ flux maps  by collapsing the datacubes over a fixed velocity range, depending on the target, for both datasets. In Table~\ref{tab:hz} we have reported the properties of the ALMA datasets and \cii\ images of the three sources. 

The two datesets have the same noise level for HZ1 and HZ2, while the sensitivity of the  ALPINE data for HZ4 is 2.5 times higher than the rms of the old observations. For HZ4, we have thus split the 2017.1.00428.L, i.e. ALPINE, dataset in six ($\approx2.5^2$) parts and generated the \cii\ datacube from one of the various sub-datasets in order to obtain a final \cii\ image with a noise level as high as that of the  2012.1.00523.S program (see Table~\ref{tab:hz}). 

By adopting the same prescription used for the mock data (see Sect.~\ref{sec:simulations}), we have measured the integrated \cii\ flux of each galaxy in both datasets. We have also used the \verb'imfit' task in \verb'CASA' to estimate the deconvolved size of \cii\ emission. The integrated flux  (Table~\ref{tab:hz}) and source extension estimates are in agreement within the error with those reported by \citet{Capak:2015}, \citet{Bethermin:2020}, and \citet{Fujimoto:2020}. 

\begin{table}
	\caption{Properties of the ALMA datasets and \cii\ images of HZ1, HZ3, and HZ4.}\label{tab:hz}

\centering
	\begin{tabular}{lcc} % four columns, alignment for each
		\hline
		     &  2017.1.00428.L & 2012.1.00523.S  \\
		\hline
		\hline
		 & \multicolumn{2}{c}{HZ1}\\
		beam & 0.88\arcsec$\times$0.80\arcsec & 0.76\arcsec$\times$0.52\arcsec \\
		$\sigma_{16}^{(a)}$ [mJy beam$^{-1}$] & 0.29 & 0.31 \\
		I$_{\rm [CII]}$ [Jy km s$^{-1}$] & $0.58\pm0.08$ & $0.36\pm0.09$ \\
		SNR$_{\rm peak}^{(b)}$ & 10.3 & 7.7 \\
		\hline
		 & \multicolumn{2}{c}{HZ3}\\
		beam & 1.32\arcsec$\times$0.99\arcsec & 0.76\arcsec$\times$0.41\arcsec \\
		$\sigma_{16}^{(a)}$ [mJy beam$^{-1}$] & 0.40 & 0.39 \\

		I$_{\rm [CII]}$ [Jy km s$^{-1}$]& $0.92\pm0.08$ & $0.67\pm0.15$ \\
		SNR$_{\rm peak}^{(b)}$ & 16.5 & 7.3 \\

		\hline
				 & \multicolumn{2}{c}{HZ4}\\
		beam & 0.96\arcsec$\times$0.81\arcsec & 0.88\arcsec$\times$0.49\arcsec \\	
		$\sigma_{16}^{(a)}$ [mJy beam$^{-1}$] & 0.24  & 0.60 \\
		& (0.55)$^{(c)}$ & \\
		I$_{\rm [CII]}$ [Jy km s$^{-1}$] & $0.89\pm0.05$ & $0.64\pm0.16$ \\
		 & \quad($0.87\pm0.18$)$^{(c)}$ &  \\
		SNR$_{\rm peak}^{(b)}$ & 18.1  & 7.2 \\
		 & \quad(9.2)$^{(c)}$ & \\
		\hline
		\hline
		\end{tabular}
	
	{\bf Notes}: $^{(a)}$ rms measured in a spectral channel of 16 \kms; $^{(b)}$ signal-to-noise ratio of the integrated \cii\ map defined as the ratio between the peak and the noise level; $^{(c)}$ rms, integrated flux measured, and SNR from the cropped dataset 2017.1.00428.L of HZ4.
	
\end{table}

\section{MS0451-H}\label{app:appb}

MS0451-H is a lensed galaxy at $z = 6.703\pm0.001$ with SFR~$= 0.4$~\sfr and magnification factor $\mu=100\pm20$ \citep{Knudsen:2016}. 
ALMA observations of this galaxy  were carried out in Cycle 2 by using a compact array configuration, leading to a natural resolution of $1.6\arcsec\times0.9\arcsec$. The data has been  discussed in \cite{Knudsen:2016} who reported an upper limit on \cii\ luminosity of $3\times10^5$~\lsun, i.e. 15 times lower than what expected from the local \lcii-SFR relation. 
By taking into account the intrinsic scatter of the \lcii-SFR relation estimated  by \cite{Carniani:2018} for the high-$z$ galaxies, $\sigma=0.48$~dex,  the upper limit deviates from the local relation by 2.5$\sigma$.

\begin{figure}
	\includegraphics[width=\columnwidth]{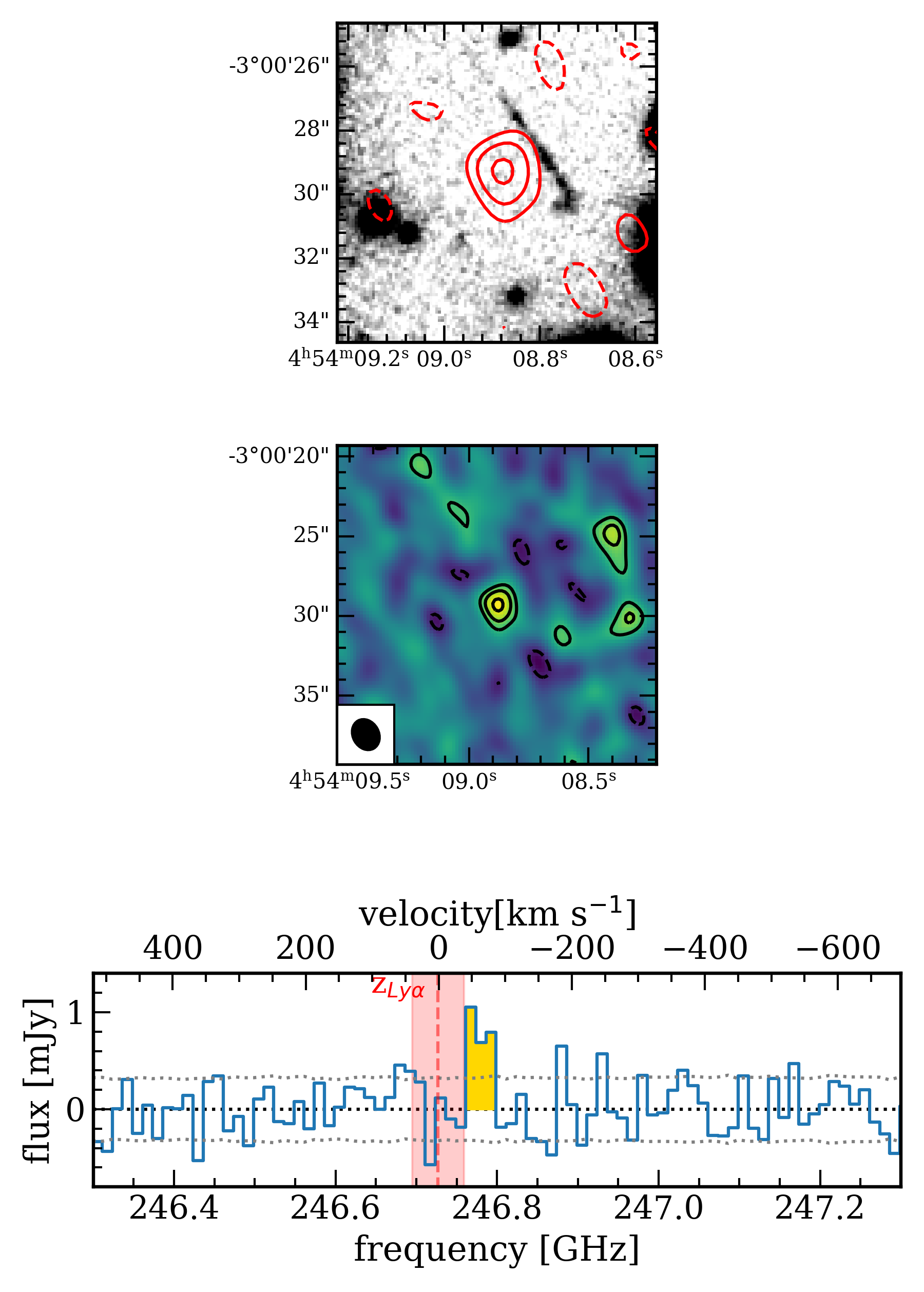}
    \caption{\emph{Top}: HST/WFC3 F110W thumbnail image of MS0451-H arc. Red contours show  $\pm$2, 3, 4$\sigma$ contours of  \cii\ emission.  \emph{ Middle}: {\it uv}-tapered \cii\ flux map obtained by integrating between -45~\kms\ and -95\kms\ with respect to the \lya\ redshift (\citealt{Knudsen:2016}). The black lines trace the $\pm$2, 3, 4$\sigma$ contours, where 1$\sigma$ level is  9~mJy~\kms. ALMA beam has been reported in the bottom-left corner. \emph{Bottom}:  spectrum of the candidate \cii\ detection, with a spectral rebinning of $\sim$16~\kms.  The vertical dashed line shows the redshift inferred from the \lya, while the grey dotted lines indicate the noise level in the ALMA cube. The gold shaded area represents the frequency range used to obtain the  \cii\ flux map shown in the middle panel. }
    \label{fig:ms0451}
\end{figure}

However, the HST/WFC3 F110W  image (i.e. rest-frame UV) shows that MS0451-H is a gravitationally  lensed arc with an extent  of  about $5\arcsec$ (Fig.~\ref{fig:ms0451}), which is $\sim3$ times larger than the ALMA beam. Therefore, the non-detection could be associated to the low surface brightness of the source. 
We have also re-analysed the ALMA data by performing different {\it uv}-tapering in order to recover the extended emission. In the {\it uv}-tapered cube with angular resolution of 2.1\arcsec\ we have found a potential \cii\ line at frequency of 246.783~GHz (Fig.~\ref{fig:ms0451}, which is consistent with the \lya\ redshift ($\Delta v = -68$~\kms) once the  intergalactic medium absorption is taken into account \citep[e.g.][]{Maiolino:2015, Pentericci:2016, Matthee:2019}.
The \cii\ emission is offset by 1\arcsec to the Est with respect to the UV emission. Since this spatial offset is slightly larger than astrometry accuracy of 0.7\arcsec, we cannot confirm if this potential emission is associate to the  arc. 

The \cii\ emission is detected in the integrated map with an \snr~$ = 4.2$
and has an integrated flux density of $42\pm10$~mJy~\kms, which
corresponds to \lcii~$= (4.8\pm1.1)\times 10^{5}(\mu/100)$~\lsun. 
We note that the error associated with such a large magnification  factor might plausibly be higher than 50\% when systematics are included 
\citep[e.g.][]{Meneghetti:2017}. Moreover, the vicinity to the critical line also 
suggests a very steep magnification gradient is present, implying a 
significant differential magnification can affect the two spatially offset regions.

If we assume that the \cii\ emission is as extended as the UV region, our simulations expect that $\sim60-80\%$ of the carbon emission is missed in current ALMA observations. We thus infer a total \cii\ luminosity of \lcii~$= 1.2-2.4\times 10^{6}(\mu/100)$~\lsun\ that is consistent with the local \lcii-SFR relation within 1$\sigma$. However, deeper ALMA observations are needed to confirm the candidate detection and to estimate the total \cii\ emission associated to this lensed galaxy.

%%%%%%%%%%%%%%%%%%%%%%%%%%%%%%%%%%%%%%%%%%%%%%%%%%

% Don't change these lines
\bsp	% typesetting comment
\label{lastpage}
\end{document}